\documentclass[prd, preprintnumbers, amsmath, amssymb, nofootinbib, aps, superscriptaddress,]{revtex4-2}

\usepackage{bm}
\usepackage{amsfonts}
\usepackage{mathrsfs}
\usepackage{graphicx}
\usepackage{amsmath}
\usepackage{float}
\usepackage{braket}

\newcommand{\beginsupplement}{%
        \setcounter{table}{0}
        \renewcommand{\thetable}{S\arabic{table}}%
        \setcounter{figure}{0}
        \renewcommand{\thefigure}{S\arabic{figure}}%
     }

\begin{document}

\newcommand{\sgn}{\operatorname{sgn}}
\newcommand{\hhat}[1]{\hat {\hat{#1}}}
\newcommand{\pslash}[1]{#1\llap{\sl/}}
\newcommand{\kslash}[1]{\rlap{\sl/}#1}
\newcommand{\lab}[1]{}
\newcommand{\iref}[2]{}
\newcommand{\sto}[1]{\begin{center} \textit{#1} \end{center}}
\newcommand{\rf}[1]{{\color{blue}[\textit{#1}]}}
\newcommand{\eml}[1]{#1}

\newcommand{\er}[1]{Eq.\eqref{#1}}
\newcommand{\df}[1]{\textbf{#1}}
\newcommand{\mdf}[1]{\pmb{#1}}
\newcommand{\ft}[1]{\footnote{#1}}
\newcommand{\n}[1]{$#1$}
\newcommand{\fals}[1]{$^\times$ #1}
\newcommand{\new}{{\color{red}$^{NEW}$ }}
\newcommand{\ci}[1]{}
\newcommand{\de}[1]{{\color{green}\underline{#1}}}
\newcommand{\ke}{\rangle}
\newcommand{\br}{\langle}
\newcommand{\lb}{\left(}
\newcommand{\rb}{\right)}
\newcommand{\lbk}{\left[}
\newcommand{\rbk}{\right]}
\newcommand{\blb}{\Big(}
\newcommand{\brb}{\Big)}
\newcommand{\nn}{\nonumber \\}
\newcommand{\p}{\partial}
\newcommand{\pd}[1]{\frac {\partial} {\partial #1}}
\newcommand{\cd}{\nabla}
\newcommand{\cc}{$>$}
\newcommand{\bqa}{\begin{eqnarray}}
\newcommand{\eqa}{\end{eqnarray}}
\newcommand{\bqe}{\begin{equation}}
\newcommand{\eqe}{\end{equation}}
\newcommand{\bay}[1]{\left(\begin{array}{#1}}
\newcommand{\eay}{\end{array}\right)}
\newcommand{\eg}{\textit{e.g.} }
\newcommand{\ie}{\textit{i.e.}, }
\newcommand{\iv}[1]{{#1}^{-1}}
\newcommand{\st}[1]{|#1\ke}
\newcommand{\at}[1]{{\Big|}_{#1}}
\newcommand{\zt}[1]{\texttt{#1}}
\newcommand{\non}{\nonumber}
\newcommand{\m}{\mu}

\def\xa{{m}}
\def\xA{{m}}
\def\xb{{\beta}}
\def\xB{{\Beta}}
\def\xd{{\delta}}
\def\xD{{\Delta}}
\def\xe{{\epsilon}}
\def\xE{{\Epsilon}}
\def\xve{{\varepsilon}}
\def\xg{{\gamma}}
\def\xG{{\Gamma}}
\def\xk{{\kappa}}
\def\xK{{\Kappa}}
\def\xl{{\lambda}}
\def\xL{{\Lambda}}
\def\xo{{\omega}}
\def\xO{{\Omega}}
\def\xvp{{\varphi}}
\def\xs{{\sigma}}
\def\xS{{\Sigma}}
\def\xt{{\theta}}
\def\xvt{{\vartheta}}
\def\xT{{\Theta}}
\def \Tr {{\rm Tr}}
\def\CA{{\cal A}}
\def\CC{{\cal C}}
\def\CD{{\cal D}}
\def\CE{{\cal E}}
\def\CF{{\cal F}}
\def\CH{{\cal H}}
\def\CJ{{\cal J}}
\def\CK{{\cal K}}
\def\CL{{\cal L}}
\def\CM{{\cal M}}
\def\CN{{\cal N}}
\def\CO{{\cal O}}
\def\CP{{\cal P}}
\def\CQ{{\cal Q}}
\def\CR{{\cal R}}
\def\CS{{\cal S}}
\def\CT{{\cal T}}
\def\CV{{\cal V}}
\def\CW{{\cal W}}
\def\CY{{\cal Y}}
\def\BC{\mathbb{C}}
\def\BR{\mathbb{R}}
\def\BZ{\mathbb{Z}}
\def\sA{\mathscr{A}}
\def\sB{\mathscr{B}}
\def\sF{\mathscr{F}}
\def\sG{\mathscr{G}}
\def\sH{\mathscr{H}}
\def\sJ{\mathscr{J}}
\def\sL{\mathscr{L}}
\def\sM{\mathscr{M}}
\def\sN{\mathscr{N}}
\def\sO{\mathscr{O}}
\def\sP{\mathscr{P}}
\def\sR{\mathscr{R}}
\def\sQ{\mathscr{Q}}
\def\sS{\mathscr{S}}
\def\sX{\mathscr{X}}

\def\slz{SL(2,Z)}
\def\slr{$SL(2,R)\times SL(2,R)$ }
\def\ads{${AdS}_5\times {S}^5$ }
\def\adst{${AdS}_3$ }
\def\sun{SU(N)}
\def\ad#1#2{{\frac \delta {\delta\sigma^{#1}} (#2)}}
\def\bqf{\bar Q_{\bar f}}
\def\nf{N_f}
\def\sunf{SU(N_f)}

\def\dcirc{{^\circ_\circ}}

\author{Morgan H. Lynch}
\email{morgan.lynch@technion.ac.il}
\affiliation{Department of Electrical Engineering,
Technion: Israel Institute of Technology, Haifa 32000, Israel}
\author{Eliahu Cohen}
\email{eliahu.cohen@biu.ac.il}
\affiliation{Faculty of Engineering and the Institute of Nanotechnology and Advanced Materials,
Bar Ilan University, Ramat Gan 5290002, Israel}
\author{Yaron Hadad}
\email{yaronhadad@gmail.com}
\affiliation{Department of Electrical Engineering,
Technion: Israel Institute of Technology, Haifa 32000, Israel}
\author{Ido Kaminer}
\email{kaminer@technion.ac.il}
\affiliation{Department of Electrical Engineering,
Technion: Israel Institute of Technology, Haifa 32000, Israel}

\title{Experimental Observation of Acceleration-Induced Thermality}
\date{\today}

\begin{abstract}
We examine the radiation emitted by high energy positrons channeled into silicon crystal samples. The positrons are modeled as semiclassical vector currents coupled to an Unruh-DeWitt detector to incorporate any local change in the energy of the positron. In the subsequent accelerated QED analysis, we discover a Larmor formula and power spectrum that are both thermalized by the acceleration. Thus, these systems explicitly exhibit thermalization of the detector energy gap at the celebrated Fulling-Davies-Unruh (FDU) temperature. Our derived power spectrum, with a nonzero energy gap, is then shown to have an excellent statistical agreement with high energy channeling experiments and also provides a method to directly measure the FDU temperature. We also investigate the Rindler horizon dynamics and confirm that the Bekenstein-Hawking area-entropy law is satisfied in these experiments. As such, we present the evidence for the first observation of acceleration-induced thermality in a non-analogue system.
\end{abstract}


\maketitle
\section{Introduction}
The machinery employed by quantum field theory in curved spacetime \cite{davies, aspects, pt, parker,parker1} to analyze radiation emission has the advantage of stripping away the minute details of the emission process and focusing entirely on the local change in energy of a radiating system. Motivated by the original analyses of Unruh \cite{unruh1} and DeWitt \cite{dewitt} to simplify the transitions in energy of quantum fields in general relativistic backgrounds by modeling them as two level systems, we return to the original use of the ``Unruh-DeWitt detector" to analyse how a radiating particle changes its energy. The fundamental insight from this perspective is that for certain deformations in the structure of spacetime, an Unruh-DeWitt detector will undergo a transition, either up or down in energy, and radiate \cite{lynch,lynch1,unruh1,dewitt,muller,matsas1,matsas3}. Remarkably, the character of this radiating system will be thermalized at a temperature defined by some characteristic inverse length scale of the spacetime or motion through it, e.g. acceleration, surface gravity, Hubble constant, or some combination thereof, \cite{unruh1,hawking,gibbons,lynch3,niayesh}. Each of these length scales will define the location of an event horizon which, via quantum fluctuations near the horizon, will emit radiation which is thermalized at that characteristic temperature.  In FRW cosmologies \cite{obadia}, signatures of this temperature may be encoded in the anisotropies in the cosmic microwave background \cite{agullo}. Analogue systems involving ``fluid black holes" \cite{dumb,silke,jeff,germain} are also capable of exploring not only the temperature, but other properties such as the entropy via correlations between Hawking pairs both inside and outside the effective horizon and even superradiant scattering. The most sought after temperature \cite{matsas4,matsas5}, due to the fact that it appears to be the most readily accessible experimentally, is the acceleration temperature $T = \frac{a}{2\pi}$. The path towards the discovery of this temperature began with the analysis of the quantum mechanical structure of the vacuum in inertial and accelerated reference frames by Fulling \cite{fulling}, the flux of radiation from the 1+1 dimensional moving mirror by Davies \cite{davies1}, and finally by the near horizon examination of Hawking radiation from black holes by Unruh \cite{unruh1}. Understanding this Fulling-Davies-Unruh (FDU) temperature has been the subject of a steadily growing community, as detailed in \cite{matsas4}, and the techniques developed to explore it have spread to other fields and fostered their grow as well, e.g. the use of Unruh-DeWitt detectors in relativistic quantum information \cite{hu}. Since the FDU temperature and the general characteristics of electromagnetic radiation are determined by the acceleration, it is through accelerated electromagnetic systems that we expect to see the first signs of accelerated thermality. 

These pursuits have culminated in an inherently thermodynamic understanding of the nature of relativistic quantum field theory in classical general relativistic backgrounds \cite{jacobson}. One of the first clues for this thermodynamic interpretation comes from the notion of black hole entropy. Bekenstein conjectured that the information content or entropy, $S$, associated with a blackhole would be proportional to the surface area, $A$, of the event horizon \cite{bekenstein}. Hawking later determined the proportionality constant to be $1/4$ and thus gave rise to the Bekenstein-Hawking area-entropy law \cite{hawking}, $S = \frac{A}{4}$. What is particularly interesting about this expression is that it specifically depends on the Planck area, $\ell^{2}_{p}$. This area-entropy law is applicable to more than just black hole and is, in fact, a general property of systems with horizons. In particular, the Rindler horizon associated with acceleration will also obey this law \cite{satz}. Consequently, given a system with sufficient acceleration, thermality can be verified and explored by both the presence of a well defined FDU temperature as well as the Rindler horizon dynamics. In fact, they both provide an independent confirmation of the presence of acceleration-induced thermality. 	  

Here we employ a spacetime formulation of \textit{accelerated quantum electrodynamics} (AQED) \cite{lynch,lynch1,lynch4}, via the use of a uniformly accelerated Unruh-DeWitt detector \cite{unruh1,dewitt} and apply it to high energy channeling radiation \cite{wistisen}. The result of this analysis is a statistically significant indication that \textit{the FDU temperature has finally been observed, in a non-analogue experiment.} We demonstrate that a nonzero Unruh-DeWitt detector energy gap, which encodes local changes in energy of the radiating electron or positron, will not only be thermalized at the FDU temperature but also provide a significantly better explanation of high energy channeling radiation than conventional models. We present the AQED response function and compute the power radiated by, and photon power spectrum of, a uniformly accelerated charge current. To compare with the channeling radiation experimental data, we assume an energy gap comprised of a Taylor series in the photon frequency and an acceleration profile based on radiative energy loss. A chi-squared analysis is shown to strongly favor the presence of a non-zero energy gap and therefore provides strong evidence for the observation of thermality at the FDU temperature. We then provide an independent verification of thermality by analyzing the Rindler horizon dynamics. We find the Bekenstein-Hawking area-entropy law is satisfied via its convergence to 4$\ell_{p}^{2}$. Here and throughout, we use natural units $\hbar = c = k_{B} = G = 1$.

\section{The AQED Response Function and Thermalized Observables}

The calculation presented here uses an AQED approach for computing the power radiated from first principles. To this end we make use of the current interaction \cite{peskin}, $\hat{S}_{I} = \int d^{4}x \hat{j}_{\m}(x)\hat{A}^{\m}(x)$, where $\hat{A}^{\m}(x)$ is the second quantized photon field operator and $\hat{j}_{\m}(x)$ is the electron current operator. If we couple an Unruh-DeWitt detector to the current it takes the form $\hat{j}_{\m}(x) = u_{\m}\hat{q}(\tau)\delta^{3}(x-x_{tr}(\tau))$. Here $u_{\m}$ is the four-velocity of the electron and $x_{tr}(\tau)$ is the trajectory of the electron parametrized by its proper time. The monopole moment operator \cite{dewitt} is Heisenberg evolved via $\hat{q}(\tau) = e^{i\hat{H}\tau}\hat{q}(0)e^{-i\hat{H} \tau}$ and the charge of the electron is defined by $q=\bra{E_{f}}\hat{q}(0)\ket{E_{i}}$, with $\ket{E_{i}}$ and $\ket{E_{f}}$ being the initial and final electron energy respectively. We must also note that our monopole moment operator creates states of definite momentum. As such for energy gaps comprised of a continuum of frequencies, such as the that we will analyze in this work, our current should be thought of as coupled to a continuum of detectors; one for each frequency. This also will restrict our working regime to first order in perturbation theory. Then, by analyzing the amplitude, $\mathcal{A} = i\bra{\mathbf{k}}\otimes \bra{E_{f}}\hat{S}_{i}\ket{E_{i}}\otimes \ket{0}$, for the electron current to undergo a transition and emit one photon of momentum $\mathbf{k}$, we can compute the emission rate via the relativistic analogue of Fermi's golden rule. As such, the AQED response function, see Sec. B of the supplementary and \cite{davies}, is given by
\bqa
\Gamma = q^{2} \int d\xi e^{-i\Delta E \xi}  U_{\m \nu}[x',x]G^{\m \nu}[x',x].
\label{response}
\eqa
Here we see the standard Fourier transform of the Wightman function, $G^{\nu \m}[x',x]$, but with indices which, for photons, represent the sum of polarization four-vectors, $\sum_{i,j}\epsilon_{i}^{\mu}\epsilon_{j}^{\ast\nu}$, which contract with the four-velocity product, $U_{\m \nu} = u_{\m}u_{\nu}$, to couple the motion to the allowed emission directions. This is nothing more than the byproduct of the standard $v\cdot A$ coupling that one typically sees in electron-photon systems. Recalling the Wightman function is given by the vacuum-to-vacuum two-point function $G^{\nu \m}[x',x] = \bra{0} \hat{A}^{\dagger \nu}(x')\hat{A}^{\m}(x)\ket{0}$, we will have
\bqa
G^{\m \nu}[x',x] = \frac{1}{(2 \pi)^{3}}\frac{1}{2} \int \frac{d^{3}k}{\omega}\sum_{i,j}\epsilon_{i}^{\m} \epsilon_{j}^{\ast\nu}e^{i(\mathbf{k}\cdot \Delta \mathbf{x} - \omega (t'-t))}.
\eqa
Since we wish to compute the power radiated, $\mathcal{S}$, we must also include an additional factor of frequency in the above Wightman function \cite{matsas1,lynch1,matsas3}, as $\mathcal{S} = \int \frac{d\Gamma}{d\omega}\omega d\omega$. When we compute the power radiated by an accelerated electron we should expect to recover the Larmor formula. As we shall see, this is indeed the case. For a trajectory with constant proper acceleration $a$, parametrized by the proper time $\tau$, we will have hyperbolic motion with four-velocity $u_{\m} = (\cosh{(a\tau)},0,0,\sinh{(a\tau)})$ \cite{matsas4}. Using this trajectory, we arrive at a rather striking result. It is expected that we should obtain the Larmor formula but there is an intermediate result; the fact we have uniformly accelerated motion implies that we should also find signatures of thermality. Explicit computation of the power, see Sec. C of the supplementary, yields
\bqa
\mathcal{S} = \frac{2}{3}\alpha a^{2} \frac{1}{1+e^{2\pi \Delta E/a}}.
\label{larmor}
\eqa     
It is indeed surprising that we have obtained the Larmor formula that is thermalized at the FDU temperature. We must also point out the change in statistics from bosonic to fermionic. This is characteristic of accelerated thermal observables where the statistics depends on, for example, powers of frequency $\omega$ in the computation, number of particles emitted, and/or the dimensionality of the system \cite{matsas1,lynch,lynch1,takagi}. To recover the standard Larmor formula we must compute the total power by summing over emission and absorption of zero energy Rindler photons \cite{matsas5,brem1,weinberg,pauri}, i.e. $ \lim_{\Delta E \to 0} \mathcal{S}(\Delta E) + \mathcal{S}(-\Delta E)=\frac{2}{3}\alpha a^{2}$. This is also corroborated by the notion that classical radiation sources can be viewed as ``gapless" limit of an Unruh-DeWitt detector \cite{gapless}. To compare the AQED theory with the experimental data, we also present the total power radiated per unit frequency, see the supplementary Eqs. (S41) and (S42),
\bqa
\frac{d \mathcal{S}}{d \omega} &=& -i \frac{4}{3}\alpha \frac{ \omega^2}{a} \lbk  \delta H^{(2)}_{\frac{2i\Delta E}{a}}\lb - \frac{2i \omega \gamma}{a} \rb  -\frac{1}{2} \lb  H^{(2)}_{\frac{2i\Delta E}{a} -2}\lb - \frac{2i\omega  \gamma}{a} \rb +  H^{(2)}_{\frac{2i\Delta E}{a} +2}\lb - \frac{2i\omega  \gamma}{a} \rb \rb \rbk \non \\
&\times & \lbk 1+ e^{2\pi \Delta E/a} \rbk. 
\label{spectrum}
\eqa
Here, we have defined the relativistic boost parameter $\delta = 2 \gamma^{2}-1$ and see that the power spectrum is comprised of Hankel function of the second kind, $H^{(2)}_{\ell}(x)$ \cite{stegun}. We also made the presence of thermality more apparent by making use of the following identity, $H^{(2)}_{\ell}(x) = e^{i\ell \pi}H^{(2)}_{-\ell}(x)$. The implication of this property of Hankel functions of the second kind is the manifest detailed balance at thermal equilibrium of the power spectrum by rigorous mathematical identity. The exponent produced by the change in sign of the Hankel index is precisely the Boltzmann factor comprised of the Unruh-DeWitt detector energy gap thermalized at the temperature $T = \frac{a}{2\pi}$. As such, we are led to the conclusion that \textit{systems described by this power spectrum imply the experimental observations of thermality at the FDU temperature.} 

This thermal phenomenon, commonly referred to as the Unruh effect \cite{unruh1}, has had a considerable amount of effort dedicated to its study as well as a detailed exploration of potential experimental settings which could measure it \cite{matsas4}. The main difficulty with measuring such an effect is the vanishingly small energy scale set by the acceleration when compared to the scale of the energy gap. Broadly speaking, the ability to probe the Unruh effect necessitates $|\Delta E| \sim \frac{a}{2\pi}$ so that the thermal distribution can be explored. The difficulty then lies in bringing the two energy scales together; finding acceleration scales that reach the energy gap from below or energy gaps that can reach the small acceleration scale from above. If both a small energy gap and a large acceleration scale can be found in an experimental system, then this would provide the best chance of measuring the Unruh effect. Through this logic, high acceleration scales coupled to small energy gaps, we apply the above power spectrum, Eq. (\ref{spectrum}), to the recent channeling radiation experiment in aligned crystals \cite{wistisen}. There we have LHC scale energetic positrons channeled into single crystal silicon (large acceleration) sensitive to the channeling oscillation \cite{channel}, recoil/radiation reaction \cite{dirac,landau,dipiazza,yaron,yarden}, as well as other processes in the comoving frame (small energy gap). Note, the use of Unruh-DeWitt detectors has also been successfully applied to exploring recoil/radiation reaction in accelerated Cherenkov systems \cite{lynch4}. As we shall see, it appears to be the case that channeling radiation not only provides a setting to measure radiation reaction, but may finally enable a system to explore the Unruh effect experimentally.

\section{Incorporating Radiation Reaction}

Understanding the nature of recoil due to photon emission from an accelerated charge has been one of the longest standing problems in physics; the problem of radiation reaction. The problem itself is typically examined by including recoil, based on the Larmor formula, in the Lorentz force, yielding the Lorentz-Abraham-Dirac (LAD) equation \cite{dirac}. However, this formalism is plagued by run away solutions which have yet to be tamed. Approximations to this equation were soon discovered \cite{landau} and subsequently solved \cite{dipiazza, yaron}. Despite this progress, the problem of run away solutions still persists \cite{yarden}. Perhaps it may be easier to incorporate recoil not at the level of the Lorentz force, but rather incorporate it into the computation of an observable using an Unruh-DeWitt detector. 

To get a better understanding of how to incorporate recoil, let us first examine a much simpler scenario, that of Cherenkov radiation. The reason for this is because the quantum recoil, or radiation reaction, correction to Cherenkov radiation is already well known. When we include an Unruh-DeWitt detector into the Cherenkov regime, we obtain the following Frank-Tamm formula \cite{lynch4},

\bqe
\frac{d \Gamma}{d \omega} =\alpha \beta  \lbk  1-\lbk  \frac{1}{n\beta}+\frac{\Delta E}{n\beta \omega \gamma} \rbk^{2} \rbk.
\eqe

The presence of the energy gap in the expression yields the anomalous Doppler effect \cite{frolov}. The interesting thing is that when we set the energy gap to $\Delta E  =   \frac{(n^{2}-1)\omega^{2}}{2m}$, then we reproduce the quantum recoil correction to Cherenkov radiation identically \cite{sokolov}. The next question is where did our energy gap, which gave us the recoil correction, come from? If we consider an electron with a renormalized rest mass \cite{tsytovich} comprised of the electron rest mass and the emitted photons energy $E_{i} = \sqrt{m^2+\omega^2}$ and a final state electron energy comprised of a bare electron and recoil momentum $k=n\omega$ produced by the photon emission, $E_{f} = \sqrt{m^2+(n\omega)^2}$, then the difference in energy is given by $\Delta E = E_{f} - E_{i} \approx \frac{(n^{2}-1)\omega^{2}}{2m}$. It is this physical interpretation, backed by mass renormalization in the Cherenkov regime, that we will utilize in the Larmor setting for the inclusion of recoil, i.e. radiation reaction.

Having successfully formulated a theory of recoil in Cherenkov emission using an Unruh-DeWitt detector, we now turn to the inclusion of recoil into the Larmor formula. Given that we know the inclusion of recoil is tantamount to including a recoil kinetic energy, which is process independent, we then look for a complete derivation of the Larmor formula, within an Unruh framework, and include an energy gap $\Delta E \sim \frac{\omega^{2}}{2m}$. This form of recoil is also present when looking at non-relativistic Unruh-DeWitt detectors with a quantized center of mass \cite{kempf}. The incorporation of a quantized center of mass also gives rise to the possibility of exploring light matter interactions which go beyond the Unruh-DeWitt detector approximation \cite{ed2}. By incorporating this recoil into our response function we can compute, via a series expansion in $\omega$, the total power radiated. The first term gives the Larmor formula, $\mathcal{S}_{0} = \frac{2}{3}\alpha a^{2}$. The second term gives the 1st order quantum recoil correction, $\mathcal{S}_{1} = -\frac{8\alpha}{ m} a^{2}T_{FDU}$, see Sec. D of the supplementary. Combining the two gives us our quantum corrected Larmor formula,
\bqe
\mathcal{S}=\frac{2}{3}\alpha a^{2}\lbk 1-\frac{12}{m}T_{FDU}\rbk.
\eqe

The functional dependence on the temperature as well as the sign are also in complete agreement with the first order correction to the massless scalar case \cite{hu1}. We can also compute the radiation reaction force for each term. If we recall the work done by the radiation reaction force is equal to the total energy radiated, we have $\int F^{rr} dx = -\int \mathcal{S} dt $. From this we find our radiation reaction forces to be $F^{rr}_{0} = \frac{2}{3}\alpha J$ and $F^{rr}_{1} = -\frac{8\alpha}{\pi m} J a$. Thus we find the 1st order quantum LAD equation to be,

\bqe
m\frac{du^{\m}}{ds} = qF^{\m\nu}u_{\nu} + \frac{2}{3}\alpha \lbk 1-\frac{24}{ m} T_{FDU} \rbk \lbk J^{\m}+a^2u^{\m}  \rbk.
\eqe
It is interesting to note that the quantum recoil correction seems to renormalize the fine structure constant by the acceleration $\alpha \rightarrow \alpha\lbk 1-\frac{24}{ m} T_{FDU} \rbk $. It is with the above considerations, namely the inclusion of $\Delta E \sim \frac{\omega^{2}}{2m}$ in our energy gap, that we may investigate recoil or radiation reaction in accelerated systems, see Sec. J of the supplementary for more detail. The existence of such a term in the energy gap of our power spectrum, Eqn. (\ref{spectrum}), when compared to the channeling radiation data set \cite{wistisen}, would indeed confirm the presence of recoil. 

\section{Measuring the Fulling-Davies-Unruh Temperature}
A recent proposal was made by Cozzella et al. which outlined a method of measuring the FDU temperature directly from a data set \cite{matsas5}. In their analysis, a longitudinally accelerated electron is subjected to a comoving cyclotron oscillation in the transverse plane and they compute the photon emission spectrum per unit transverse momentum. The analysis was carried out in both Minkowski space, i.e. the lab frame, as well as Rindler space, i.e. the comoving frame. General covariance necessitates that the emission rates be identical and in order to accomplish this, the Rindler frame computation must take into account the emission and absorption of Rindler photons from a background thermal distribution at the FDU temperature; In other words, the Unruh effect is mandatory to render the process covariant \cite{matsas3}. Most importantly, the temperature of the Rindler bath can be left arbitrary and the resultant observable can be matched to the data set by finding the best fit temperature; thereby providing a method to directly measure the FDU temperature.

What is particularly intriguing about this proposal is that it closely resembles the dynamics of the channeling radiation experiment \cite{wistisen}. There, a positron undergoes a longitudinal acceleration and oscillates transversely, i.e. a one dimensional oscillation. The only real difference is the fact that channeling experiment is ultra relativistic. As such, if we are able to reproduce our thermalized power spectrum via a Rindler frame computation, we can not only gain insight into the nature of the processes present in the comoving frame, but then also leave the Rindler temperature term arbitrary to yield a more general expression to be applied to the channeling data set. Should we find that the power spectrum accurately describes the data, we can then provide a direct measurement of the temperature, thereby confirming the presence of a thermalized Rindler bath and thus the Unruh effect itself.

In order to analyze the Rindler space emission rate, we will utilize the formalism developed in \cite{matsas5,kolekar} we must first consider the Rindler coordinate system $(\tau, \xi,x,y)$, with line element given by $ds^{2} = e^{2a\xi}(d\tau^{2} - d\xi^{2})-dx_{\perp}^{2}$. The two Rindler coordinates $(\tau,\xi)$ are related to the laboratory time and z coordinate via; $t = (e^{a\xi}/a)\sinh{(a\tau)}$ and $z = (e^{a\xi}/a)\cosh{(a\tau)}$. To describe the channeling oscillation we adopt a non relativistic comoving transverse oscillation, in the Rindler coordinate system, described by the four velocity $u^{\m} = (1,0,v_{0}\cos{(\Omega \tau)},0)$. Here, the amplitude of the channeling oscillation velocity $v_{0} = A \Omega$, with $A$ being the amplitude of the oscillation and $\Omega$ the channeling oscillation frequency in the comoving frame.

It has been well established that photon emission into Minkowksi space corresponds to both the emission of Rindler photons from and absorption of Rindler photons into the thermalized Rindler bath, due to the Unruh effect, in the comoving frame \cite{matsas4,matsas5,brem1,kolekar}. This is accomplished by weighting the probability of absorption of a Rindler photon by a thermal distribution and the emission probability by a thermal distribution plus 1, i.e. $\mathcal{P}_{abs} \sim |\mathcal{A}_{abs}|^{2}(1/(e^{\omega_{r}/T}-1))$ and $\mathcal{P}_{emi} \sim |\mathcal{A}_{abs}|^{2}(1+1/(e^{\omega_{r}/T}-1))$. Note this also makes use of the fact that $|\mathcal{A}_{abs}|^{2} = |\mathcal{A}_{emi}|^{2}$. The temperature of the background thermal bath, $T$, is kept arbitrary. The total Rindler emission rate is then given by, see Sec. G of the supplementary,  
\bqa
\Gamma_{r} &=& \int_{\infty}^{\infty} d^{2}k_{\perp} \int_{0}^{\infty} d\omega_{r} \frac{\lbk |\mathcal{A}^{1}_{abs}|^{2} + |\mathcal{A}^{2}_{abs}|^{2}   \rbk}{\Delta \tau} \coth{(\omega_{r}/(2T))}.
\eqa
Here the two terms correspond to summing over both Rindler photon polarizations. Each polarization is described by the mode functions, $f(x) = K_{i\omega_{r}/a}{\lb \frac{k_{\perp}}{a}e^{a\xi} \rb}e^{i(k_{\perp}\cdot x_{\perp}-\omega_{r}\tau)}$. Note, these photons are comprised of plane waves transverse to the direction of acceleration and a K2 Bessel function along the acceleration axis. The transverse plane waves are characterized by their momenta $k_{\perp}$ and the K2 Bessel is characterized by the Rindler frequency $\omega_{r}$. What is important to note at this point is that the contribution of the background thermal distribution is contained in the factor, $\coth{(\omega_{r}/(2T))}$. When analyzing the case of Rindler photon emission and absorption due to our transverse channeling oscillation, we then obtain the following total emission spectrum per transverse momentum,
\bqa
\frac{d\Gamma_{r}}{d^{2}k_{\perp}} &=& -\frac{i \alpha}{2\pi a} \sinh{(\pi \omega_{r}/a)}  \coth{(\omega_{r}/(2T))} e^{\frac{\pi \omega_{r}}{a}} \non \\
& \times & \lbk    \delta  H^{(2)}_{\frac{2i\omega_{r}}{a}}{\lb  \frac{-2ik_{\perp} \gamma }{a}\rb} -\frac{1}{2} \lb  H^{(2)}_{\frac{2i\omega_{r}}{a} +2}{\lb\frac{ -2ik_{\perp} \gamma }{a}\rb} +H^{(2)}_{\frac{2i\omega_{r}}{a} -2}{\lb \frac{-2ik_{\perp} \gamma }{a}\rb}	\rb \rbk.
\eqa
Here we see that both the thermal contribution as well as the indices of the Hankel functions are determined by $\omega_{r}$. For our transverse channeling oscillation, this sets $\omega_{r} = k_{x}A\Omega$. We must also comment on the fact that we have two new thermal terms. First, we have an additional factor of $\sinh{(\pi \omega_{r}/a)}$ due to the Rindler mode, $K_{n}(x)$, normalization. Second, we have a factor $e^{\frac{\pi \omega_{r}}{a}}$ which comes from the mode transformation $K_{i\omega_{r}/a}(k_{\perp}/a) \sim e^{\omega_{r}/T_{FDU}}H_{i\omega_{r}/a}(-ik_{\perp}/a)$. Now, by comparing this expression to that of the case of Minkowski photon emission we verify that the expressions match when the temperature of the Rindler bath is set to the FDU temperature. When analyzing photon emission using an Unruh-DeWitt detector in Minkowski space, we must sum over both the excitation and deexcitation of the detector \cite{gapless}, i.e. $\frac{d\Gamma_{tot}}{d^{2}k_{\perp}}=\frac{d\Gamma_{\Delta E}}{d^{2}k_{\perp}}+\frac{d\Gamma_{-\Delta E}}{d^{2}k_{\perp}}$. The resulting Minkowski emission rate yields,

\bqa
\Gamma_{m	} &=& \frac{1}{\Delta \tau}\int d^{2}k_{\perp}dk_{z} \lbk |A^{m}_{\uparrow}|^{2} +|A^{m}_{\downarrow}|^{2} \rbk \non \\
&=& \frac{1}{\Delta \tau}\int d^{2}k_{\perp}dk_{z} |A^{m}_{\uparrow}|^{2} \lbk 1+e^{\Delta E/T_{FDU}}  \rbk.
\eqa 

Note here we made use of the detailed balance relationship, $\Gamma^{m}_{\downarrow} = \Gamma^{m}_{\uparrow}e^{\Delta E/T_{FDU}}$, demonstrating thermal equilibrium at the FDU temperature. What is important to note is that the presence of thermality is entirely encoded in the Boltzmann-like factor $ 1+e^{\Delta E/T_{FDU}}$, and does not depend on the process specific Minkowski amplitudes $A^{m}$. Of course, we know this relation is also manifested in the emission rate via our Hankel modes, $H^{(2)}_{\frac{-2i\Delta E}{a}}\lb x\rb = e^{2\pi \Delta E/a}H^{(2)}_{\frac{2i\Delta E}{a}}\lb x \rb $. As such, our transverse Minkowski emission spectrum is given by,

\bqa
\frac{d\Gamma_{m}}{d^{2}k_{\perp}}&=& \frac{-i\alpha}{4\pi a}\lbk  \delta H^{(2)}_{\frac{2i\Delta E}{a}}\lb - \frac{2i k_{\perp} \gamma}{a} \rb  -\frac{1}{2} \lb  H^{(2)}_{\frac{2i\Delta E}{a} -2}\lb - \frac{2ik_{\perp} \gamma}{a} \rb +  H^{(2)}_{\frac{2i\Delta E}{a} +2}\lb - \frac{2ik_{\perp} \gamma}{a} \rb \rb \rbk \non \\
&\times & \lbk 1+ e^{2\pi \Delta E/a} \rbk. 
\eqa

Now, given the fact that both Minkowski and Rindler emission rates describe precisely the same physics, the rates are then equal $\frac{d\Gamma_{m}}{d^{2}k_{\perp}} = \frac{d\Gamma_{r}}{d^{2}k_{\perp}}$. This then yields our thermal transformation function,

\bqa
\frac{d\Gamma_{m}}{d^{2}k_{\perp}} &=& \frac{d\Gamma_{r}}{d^{2}k_{\perp}} \non \\
\frac{1}{2}\lbk 1+ e^{\Delta E/T_{FDU}} \rbk &=& 
 \sinh{(\pi \omega_{r}/a)}  \coth{(\omega_{r}/(2T))} e^{\frac{\pi \omega_{r}}{a}}\label{conv}
\eqa

We see that the above emission spectra, computed in both Rindler space with a background thermal bath and in vacuum Minkowski space, are indeed identical provided we identify our energy gap with Rindler frequency, $\Delta E = \omega_{r}$ and the background temperature with the FDU temperature $T = \frac{a}{2\pi}$. Functionally, what we have done was utilized the covariance of the photon emission between the Minkowski and Rindler frames in order to ``track" the presence of thermality. We have detailed balance from the Minkowski side, along with the photon exchange with the Rindler bath on the Rindler side scaled by the mode normalization. This relationship is process independent and demonstrates the fundamental nature of how the Unruh effect is mandatory for the self consistency of quantum field theory \cite{matsas3}; \textit{detailed balance in Minkowski space} $\leftrightarrow$ \textit{photon exchange with Rindler bath in Rindler space}. 

Now, based on the thermal identity, Eq. (\ref{conv}), we can write our full Minkowski power spectrum with the temperature of the Rindler bath left arbitrary. In short, we are converting the detailed balance relation back into the Rindler thermal bath. This will allow us to directly measure the temperature from the data set. Hence,  
\bqa
\frac{d \mathcal{S}}{d \omega} &=& -i \frac{4}{3}\alpha \frac{ \omega^2}{a} \lbk  \delta H^{(2)}_{\frac{2i\Delta E}{a}}\lb - \frac{2i \omega \gamma}{a} \rb  -\frac{1}{2} \lb  H^{(2)}_{\frac{2i\Delta E}{a} -2}\lb - \frac{2i\omega  \gamma}{a} \rb +  H^{(2)}_{\frac{2i\Delta E}{a} +2}\lb - \frac{2i\omega  \gamma}{a} \rb \rb \rbk \non \\
&\times & \sinh{(\pi \Delta E/a)}  \coth{(\Delta E/(2T))} e^{\frac{\pi \Delta E}{a}}. \label{temp}
\eqa

Provided the thermalized power spectrum, Eq. (\ref{spectrum}), accurately describes the channeling data, we can then use the above expression to fit the temperature. A successful measurement of the FDU temperature would confirm the presence of a thermal bath in the Rindler frame and thus the Unruh effect; thereby realizing the proposal of Cozzella et al. \cite{matsas5}.
\section{The Bekenstein-Hawking Area-Entropy Law}

Given a system which is manifestly thermalized by the acceleration also prompts an additional analysis of the horizon thermodynamics. Building upon the work of Bekenstein, who reasoned that the entropy of a black hole would be proportional to the area, Hawking, upon the discovery of black hole evaporation, was able to fix the proportionality constant to 1/4. This gave birth to the Bekenstein-Hawking entropy-area law \cite{bekenstein,hawking}, $S = \frac{A}{4}$. Measuring such a quantity from an astrophysical black hole seems well beyond our current experimental capabilities. However, systems which are thermalized by the acceleration obey the same area-entropy law due to the Unruh effect \cite{satz}. As such, for highly accelerated systems, one can directly test the hypothesis of Bekenstein and Hawking.

In the proper frame, the change in the horizon area is determined by the amount of energy radiated by each positron, $\Delta \tilde{E}$, into the horizon. This quantity we will extract from the actual data via $\Delta \tilde{E}=\frac{4c}{3x_{0}}\int\frac{d\mathcal{S}_{data}}{d\omega}d\omega dt$, see Sec. A of the supplementary material for more details. The corresponding change in horizon area \cite{satz} is given by $\Delta A = \frac{G c^{5}}{\hbar} \frac{8 \pi m^{3} \Delta \tilde{E}}{E_{i}^{3} a }$, see Sec. I of the supplementary. From the first law of thermodynamics, we can also determine the entropy difference in the proper frame based on the difference in positron energy in the lab frame, $\Delta S = \frac{c^{8}}{\hbar^{2}} \frac{\pi m^3  }{ a} \lbk \frac{1}{(E_{i} - \Delta \tilde{E})^{2}}-\frac{1}{E^{2}_{i}} \rbk$. Note, we reintroduced all fundamental constants into the above expressions for the area and entropy change so we can resolve the presence of the Planck area, $\ell_{p}^{2} = \frac{G\hbar}{c^{3}}$. Finally, if the hypothesis of Bekenstein and Hawking is true, then we will have $\frac{\Delta A}{\Delta S} = 4\ell^{2}_{p}$. As such, we will integrate the data in order to compute the following area to entropy ratio, see Sec. I of the supplementary,

\bqa
\frac{\Delta A}{\Delta S} = \ell^{2}_{p}\frac{8 \Delta \tilde{E}}{E^{3}_{i}}\lbk \frac{1}{(E_{i} - \Delta\tilde{E})^{2}} -\frac{1}{E^{2}_{i}} \rbk^{-1}. \label{bh}
\eqa 

When the system thermalizes, it is expected that the above expression, with the appropriate integration over the data, will converge to $4\ell_{p}^{2}$. What is required to confirm the presence of thermality is that this expression satisfy the relation $S = A/4$, even with $\Delta E \neq 0$. Conceptually what this means is that by the original integration of the 1st law, we fix $S_{i} = \frac{A_{i}}{4}$. This is the zeroth order ``initial condition" of the integration. Then, we must have $\Delta E$ evolve in such a way that the change in the area and entropy also obey $\Delta S = \frac{ \Delta A}{4}$. This is not always the case in fact. The $\Delta E$ that presents itself in the above expressions must come from a thermalized observable. When applied to the high energy channeling experiment, a verification of this area-entropy law will provide independent corroborating evidence for the presence of acceleration-induced thermality which would compliment the analysis based on the Minkowski and Rindler thermal power spectra.   

\section{Experimental Observation of Acceleration-Induced Thermality in Channeling Radiation}

When a highly energetic charged particle propagates in a material, it will lose energy and emit a smooth spectrum of radiation known as bremsstrahlung \cite{jackson}. This particular process is associated with scattering off individual atomic sites that have a random distribution, i.e. an amorphous crystalline structure for solids or the random distribution of atoms in the case of a liquid or gas. However, if we have a solid with some periodic crystalline structure, then we can look at motion along an axis of symmetry where the charged particle is ``channeled" and will propagate in an effective hollow wave guide, or 2-D plane between atomic layers, produced by the structure. Then, the charged particle is confined to the potential well and can oscillate back and forth transversely to its direction of propagation and therefore radiate. This process is known as channeling radiation \cite{gemmell,ulrik}. In the channeling experiment \cite{wistisen}, the data from the photon power spectrum produced by the rapidly decelerated positrons is presented and is precisely the observable computed in our analysis. There, $178.2$ GeV positrons are fired into a 3.8 mm sample of single crystal silicon aligned along the $\braket{111}$ axis. Because the positron is positively charged, it will be repelled by the positively charged atomic sites, and therefore be subject to a confining harmonic oscillator potential. The transitions between these harmonic oscillator states gives rise to the channeling radiation seen in the experiment. These transitions between states, i.e. the transverse oscillation, decouples from the longitudinal motion, i.e. the beam velocity. The energy loss produced by this photon emission then gives rise to the large accelerations necessary to bring about thermality.

If we consider a characteristic lab frame photon frequency, $\omega_{0}$, produced by the channeling radiation, we can determine the momentum scale of the associated process. This will resultant in a change in momentum imparted on the positron given by $|\Delta p| = |k_{0}| = \omega_{0}$. This momentum change occurs during the time scale of the emission process. Taking the physical size of the photon to be $\Delta x = \lambda_{0}/2$, we can then determine the emission time to be, $\Delta t = \Delta x/c = \frac{ \pi}{\omega_{0}}$. Then, recall the relativistic version of Newtons law, $f = \frac{\Delta p}{\Delta t} = ma'$, will enable us to determine the proper acceleration, $a'$, see Sec. H of the supplementary. Hence, we find our proper acceleration is given by,

\bqe
a' = \frac{\omega_{0}^{2}}{ \pi m}
\eqe

This is a proper acceleration but it is written in terms of the lab frequency. What is important to note is that when written in terms of proper quantities, the proper acceleration, $a' =\frac{\omega'^{2}_{max} \gamma^{2}}{ \pi m}$, boosts as $\gamma^{2}$. More importantly, when computing the FDU temperature for the emission, we find a recoil/radiation reaction temperature, $T_{RR}$. Hence,

\bqe
T_{RR} = \frac{\omega'^{2}_{0} \gamma^{2}}{ 2 m \pi^{2} }.
\eqe

Note, we now have an FDU temperature which depends explicitly on the recoil kinetic energy, $\omega^{2}/2m$ which is imparted on the positron by the emission. It is this radiation reaction temperature, produced by the recoil itself, that we will look for in our experimental analysis.

To compare with the experimental data we need to convert all parameters to GeV \cite{pdg,wistisen};  $E_{0} = 178.2 \,\textrm{GeV}$, and $m  = .000511 \, \textrm{GeV}$. We have an overall scale factor, $s$, to take into account detector efficiencies. To model the change in the positron's energy, we adopt a polynomial energy gap of the form $\Delta E = a_{0}+a_{1}\omega +a_{2}\omega^{2}+a_{3}\omega^{3}$; the logic being that the exact expression of the energy difference, produced by the relevant dispersion and conservation of momentum and energy relations associated with all processes present, will be amenable to a Taylor expansion in powers of the emitted photon's frequency \cite{chen}, see Sec. J of the supplementary for more details. These parameters will be used to compute the total power spectrum, $\frac{d\mathcal{S}_{tot}}{d\omega} = \frac{d\mathcal{S}(\Delta E)}{d\omega}+\frac{d\mathcal{S}(-\Delta E)}{d\omega}$ , with and without the energy gap, and then compared with the crystal data \cite{wistisen}. In principle, the $a_{0}$ term is associated with the channeling oscillation, $a_{1}$ is associated with the Rindler frequency, and the $a_{2}\omega^{2}$ term is associated with recoil or radiation reaction \cite{dirac,landau,dipiazza,yaron,yarden}. We included a frequency dependence up to an $\omega^{3}$ term to capture any further dependence beyond the recoil term since that was what the experiment \cite{wistisen} measured. The energy gap can be ``turned off" by setting it to zero.  To better match the calculated spectrum to the data we also include an overall scaling factor $s$ to take into account detector efficiencies and other systematics of the experiment, see Sec. A of the supplementary for more details. To ensure the presence of thermality we must also examine the emission lifetime. Recalling the emission rate per unit frequency is given by, $\frac{d\Gamma(\omega)}{d\omega} = \frac{d\mathcal{S}(\omega)}{d\omega}\frac{1}{\omega}$, we then invert this and integrate up to a specific frequency to determine the thermalization time for that frequency, i.e. the partial thermalization time \cite{wong}, $t(\omega) = \frac{1}{\int_{0}^{\omega}\frac{d\Gamma(\omega')}{d\omega'}d\omega'}$. Thermality neccesitates that we must have this lifetime be shorter than the travel time within the crystal. This will then require a low frequency cutoff for frequencies that do not have enough time for thermalization to take hold. We then the compute best fits of our theory, with respect to the paramaters s, Fulling-Davies-Unruh temperature $T_{FDU}$, Rindler temperature $T_{R}$, and $a_{i}$, along with the reduced chi-squared statistic for each low frequency cutoff. 

The plot of our emission lifetime and reduced chi-squared statistics for each cutoff for the 3.8 mm channeling crystal \cite{wistisen} is presented in Fig. 1. Also included is the best fit power spectra Eq. (\ref{spectrum}) along with examples of changing the Rindler bath temperature Eq. (\ref{temp}), and data \cite{wistisen}, which yielded the first statistically significant signals of accelerated thermality based on the chi-squared statistic. To verify the presence of thermality we also examine the Bekenstein-Hawking area-entropy law Eq. (\ref{bh}). It is through its convergence to $4\ell_{p}^{2}$, along with the chi-squared of order unity, that we are led to the conclusion of acceleration-induced thermality.

\begin{figure}[H]
\centering  
\includegraphics[scale=.39]{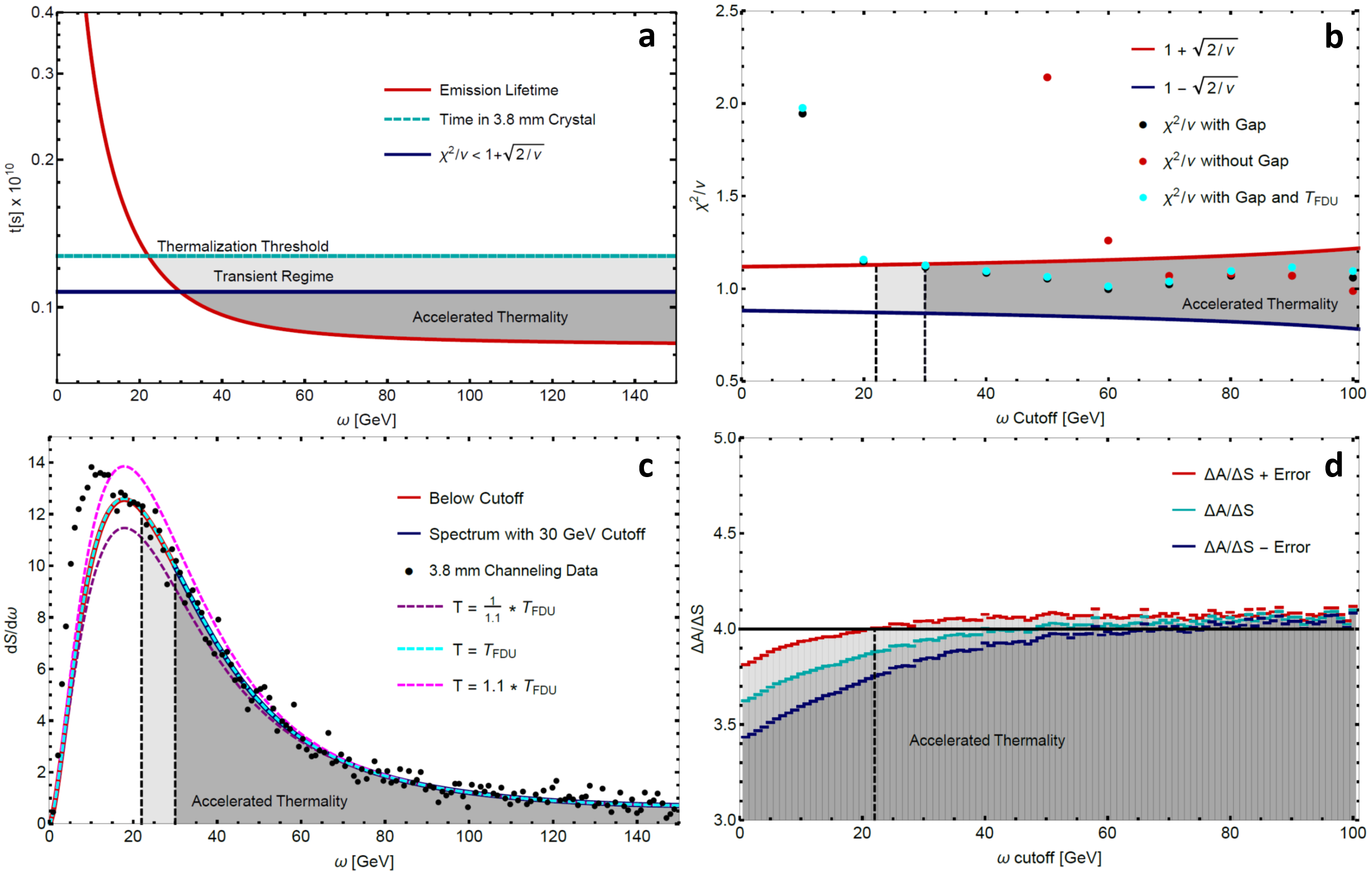} 
  \caption{\textbf{Experimental observation of acceleration-induced thermality:} In order for the Unruh effect to manifest itself, whatever process that takes place must have time to thermalize. Plot (\textbf{a}) demonstrates that the emission lifetime of the channeling experiment is shorter than the time the positron traverses the crystal sample, thereby demonstrating that the system has time to thermalize. Note the threshold energy is $\sim 22$ GeV. With the threshold of thermality exceeded, we must then look at various cutoffs beyond this threshold for a statistical signal which will confirm it. Plot (\textbf{b}) shows the chi-squared statistic for the best fit power spectrum with various low frequency cutoffs. Included are also the chi-squared for the power spectrum, Eq. (\ref{temp}), with the best fit Rindler temperatures. The channeling radiation has a $\chi^{2}/\nu <1+\sqrt{2/\nu}$ starting at 30 GeV. Moreover, the presence of a thermalized energy gap is favored over no gap. Plot (\textbf{c}) presents the power spectrum with a thermalized energy gap along with the channeling data for the first cutoff, $30$ GeV, that satisfies the chi-squared criterion. Also included is the same power spectrum with the Rindler temperature set to $T_{FDU}$, $(1.1)T_{FDU}$, and $(1/1.1)T_{FDU}$ to illustrate the sensitivity of the spectrum to the temperature. The thermalized power spectrum Eqs. (\ref{spectrum},\ref{temp}), show an excellent agreement with the data set and provide evidence for the first observation of the Unruh effect. Plot (\textbf{d}) shows the ratio of the Rindler horizon area to the entropy as a function of the low energy cutoff. For thermalized systems, this ratio is identically $4\ell_{p}^{2}$. Again, note the thermalization threshold energy of $\sim 22$ GeV; thereby providing corroborating evidence for acceleration-induced thermality.}
\end{figure}

To summarize our analysis, we found sufficient time inside the crystal for the system to be thermalized by the acceleration. The thermalization time implied a low energy cutoff of $\sim 22$ GeV for the channeling data. Performing best fits of our theory to the data, with cutoffs at multiples of 10 GeV, yielded a reduced chi-squared statistic within the 1 standard deviation threshold at 30 GeV. For this cutoff, the overall chi-squared statistic also favors a nonzero energy gap with differences, $\Delta \chi^{2} = |\chi^{2}_{gap} - \chi^{2}_{no \; gap}| = 364$. The best fit parameter for the $\omega^2$ term is $a_{2} = 846\; GeV^{-1}$, which accurately reflects the expected value of $\frac{1}{2m}\sim 978.2 \; GeV^{-1}$, thereby confirming the presence of recoil and thus radiation reaction. The best fit value for channeling frequency is $a_{0}=5.6 \; eV$, in the proper frame, which is typical for channeling oscillations \cite{ulrik}. We can also confirm the presence of the fiducial Rindler frequency term, $\omega_{r} \sim \omega A \Omega$, by the coefficient on the linear term $\omega$. For an amplitude on the order of the lattice constant for silicon, $A = 5.4 \; \AA$ and using the best fit value for the channeling frequency, $\Omega = a_{0}$, we expect the linear coefficient to be $A\Omega \sim .015$. We indeed find our best fit value to be $a_{1} = .012$, thereby confirming the the Rindler frame analysis, i.e. that Minkowski photon emission corresponds, at the very least, to the emission and absorption of Rindler photons at the Rindler frequency set by the channeling oscillation $\omega_{r} \sim A\Omega$, \cite{matsas5,kolekar}. This implies that our energy gap is given by, $\Delta E \sim \Omega + \Omega A \omega +\frac{\omega^{2}}{2m}$; the sum of a pure channeling frequency $\Omega$, the Rindler frequency $\omega_{r} = A\Omega \omega$, and the recoil term $\frac{\omega^{2}}{2m}$. 

By keeping the FDU temperature arbitrary and performing best fits for the acceleration for each power spectrum we can measure the FDU temperature, $T_{FDU}$, with the 8 cutoffs which satisfy the chi-squared criterion; $30 \; GeV - \; 100 \; GeV$. Then by fixing the acceleration to the best fit value for each cutoff we can perform the best fit for the Rindler temperature, $T_{R}$, specifically for these same cutoffs. Note the source of error for both these measurements is the standard deviation from the mean and therefore will decrease as $1/\sqrt{n}$, which for our current analysis utilized $n = 8$. For the characteristic frequency of the radiation reaction temperature, $T_{RR}$, let us use the first frequency to thermalize from the data set, i.e. $\omega_{0} = 150$ GeV. Note, this will also serve as an upper bound on the temperature. We can now compare the experimental measurements for the average FDU temperature, the average Rindler temperature, and the radiation reaction temperature. Hence,
\bqa
T_{FDU} & =& 1.80 \pm .51 \; PeV \non \\
T_{R} & =& 1.96 \pm .49 \; PeV \non\\
T_{RR} & =& 2.23 \; PeV 
\eqa

As such we find the temperature of the Rindler bath to be $T_{R} = T_{FDU}(1.09 \pm .41)$. This experimental measurement confirms the predictions of Fulling \cite{fulling}, Davies \cite{davies1}, and Unruh \cite{unruh1} and also realizes the proposal put forward by Cozzella et al. for confirming the presence of the Unruh effect by directly measuring the FDU temperature  \cite{matsas5}. The range of frequencies which set the radiation reaction temperature is also given by $\omega_{0} = 135 \; \pm \; 17.6$ GeV. The agreement of both the FDU temperature with the radiation reaction temperature also corroborates the presence of recoil. This result seems natural under framework of radiation reaction; the acceleration scale of the system appears to be set by the recoil energy of the first frequencies to thermalize, in this case $\omega_{0} \sim 150$ GeV. This immense recoil acceleration, $a \sim 5 \times 10^{39}$ $m/s^{2}$, gives rise to a temperature of $2 \times 10^{19}$ K. The incredibly high energy scales here are due to the fact that for the channeling experiment \cite{wistisen}, the Lorentz gamma is $\gamma = 3.5 \times 10^{5}$. Other prospects for future experiments could even include two photon emission processes, \cite{schutzhold}, as well as exploring finite size effects such as gauge invariance in minimal and dipole coupling interactions, \cite{ed1,ed2}.

In short, the first three terms in the Unruh-DeWitt detector energy gap all correspond to expected values. In particular the presence of an $\sim \frac{\omega^{2}}{2m}$ term confirms the presence of recoil. The theory employed in the analysis obeys detailed balance, is thermalized at the FDU temperature, and also reproduces the Larmor formula in the appropriate limit \cite{matsas5,brem1,weinberg,pauri}. Through a complimentary Rindler analysis, we confirmed the presence of a thermal bath of Rindler photons at the FDU temperature in the comoving frame \cite{matsas5,kolekar}. Then, through a completely independent thermodynamic analysis, we found the Bekenstein-Hawking area-entropy law was also satisfied starting at the same $\sim 22$ GeV threshold. It is through the chi-squared analysis of the best fit power spectrum, measurement of the FDU temperature, and confirmation of the Bekenstein-Hawking area-entropy law that we are led to the conclusion of acceleration-induced thermality.

\section{Conclusions}
The main focus of this manuscript dealt with the presence of accelerated thermality in a high energy channeling radiation experiment that measured radiation reaction \cite{wistisen}. However, there are two additional experiments that also report evidence of radiation reaction using plasma wakefield acceleration \cite{cole,poder}. There, rather than the photon power spectrum, the final state electron energy was measured. It would be an interesting avenue of research to apply this formalism to their experiments in search of accelerated thermality as well. The connection between the Unruh effect and radiation reaction has long been discussed in the literature \cite{blh1,blh2} and it appears that these, and future, systems may provide a robust experimental setting to not only explore these intriguing aspects of radiation reaction, but also investigate the intriguing nature of acceleration-induced thermality.

In this article, we employed the theory of accelerated quantum electrodynamics and used it to explore the radiation produced by uniform accelerated motion. When applied to the problem of channeling radiation we are able to incorporate a local change in energy, utilizing techniques from quantum field theory in curved spacetime, by setting the energy gap of an Unruh-DeWitt detector equal to a general polynomial, which incorporates recoil/radiation reaction, in the emitted photons frequency thereby connecting it with the Unruh effect. The presence of the Unruh effect is explored via a Rindler frame analysis which also provided a path to directly measure the FDU temperature. This work not only explores channeling radiation, in a quantitative manner, but also sheds new light on it and the Unruh effect in a manner that is backed by the experimental evidence. Moreover, by analyzing the Rindler horizon thermodynamics, we were also able to confirm the Bekenstein-Hawking area-entropy law. In conclusion, our analysis indicates that \textit{in addition to measuring radiation reaction, the recent high energy channeling radiation experiment has a significant statistical signal for the first observation of acceleration-induced thermality backed by an independent thermodynamic analysis of the Bekenstein-Hawking area-entropy law; Thereby indicating the first observation of the Unruh effect in a non-analogue system.}

\section{Acknowledgments}
We would like to thank Ulrik Uggerhoj, Niayesh Afshordi, Jorma Louko, Ted Jacobson, Alejandro Satz, Moti Segev, Jeff Steinhauer, Germain Rousseaux, Christian Nielson, Yaniv Kurman, and Alexey Gorlach for their valuable input and Tobias Wistisen for sending us the experimental data. We are also indebted to an anonymous referee for many helpful comments that helped to improve our manuscript. This work was supported by the Israel Science Foundation (ISF) grant no. 830/19 and the ERC starting grant NanoEP 851780 from the European Research Council. E.C. was supported by Grant No. FQXi-RFP-CPW-2006 from the Foundational Questions Institute and Fetzer Franklin Fund, a donor advised fund of Silicon Valley Community Foundation, the Israel Innovation Authority under Projects No. 70002 and No. 73795, the Quantum Science and Technology Program of the Israeli Council of Higher Education, and the Pazy Foundation. E.C. acknowledges helpful discussions with the Bristol reading group regarding the Unruh effect. M.H.L. was supported at the Technion by a Zuckerman fellowship.

\goodbreak

\pagebreak

	\section{Supplementary Material}

\beginsupplement

\setcounter{equation}{0}
\setcounter{figure}{0}
\setcounter{table}{0}
\setcounter{page}{1}
\makeatletter
\renewcommand{\theequation}{S\arabic{equation}}
\renewcommand{\thefigure}{S\arabic{figure}}

\subsection{Experimental Methods}
To generate our figures, the power spectrum was scaled by $s\frac{3x_{0}}{4c}$ with $x_{0} = 9.37$ cm for positrons in silicon. This scaling factor is included in our analysis since it is used by the experimental group \cite{wistisen} to take into account the detectors signal conversion process to ensure a single photon spectrum. We best fit the acceleration with a factor $\tilde{a}$   and our energy gap is given by a general polynomial of the form $\Delta E = a_{0}+a_{1}\omega +a_{2}\omega^{2}+a_{3}\omega^{3}$. We then performed a least squares best fit, with cutoffs at multiples of $10$ GeV, to obtain the values for our six parameters $s$, $\tilde{a}$, and $a_{i}$. In the case of no energy gap we merely set $\Delta E = 0$. To directly compare to the crystal data set \cite{wistisen}, our spectrum was multiplied by a factor of $\frac{3 x_{0}}{4 c} = 2.34 \times 10^{-10}$ s. This means the power spectrum we plot is $\frac{d\mathcal{S}}{d\omega} \rightarrow s\frac{3x_{0}}{4c}\frac{d\mathcal{S}}{d\omega}$. Note, here we explicitly put in the speed of light. We then plot our power spectrum with the best fit parameters for the first cutoff to satisfy the chi-squared criterion. To compute the chi-squared statistic, we evaluated our best fit power spectra at the x-value of each data point to compare the theoretical y-value to the data. The chi-squared per degree of freedom is then given by $\chi^{2}_{the}/\nu = \sum_{i}\frac{(y_{the}(x^{i}_{exp})-y^{i}_{exp})^{2}}{\sigma_{i}^{2}}$. Here $i$ labels the data points and $\sigma_{i}$ is the experimental error of each data point. The number of degrees of freedom is given by $\nu = n-p$, with $n=150$, $p_{\Delta E} = 6$, and $p_{0}=2$. Here, $p_{\Delta E}$ and $p_{0}$ are the number of fit parameters for the power spectrum with and without the energy gap respectively.

In order to analyze the area-entropy law with a low energy cutoff in the energy radiated let us define $\Delta \tilde{E}(\omega_{c},\omega_{f}) = \frac{4c}{3x_{0}}\int_{\omega_{c}}^{\omega_{f}}\frac{d\mathcal{S}_{data}}{d\omega}d\omega dt$. Here, $\omega_{c}$ and $\omega_{f}$ are the cutoff frequency and the final frequency of the emitted photons respectively. Then, we will have the energy radiated given by $\Delta \tilde{E}(\omega_{c},150)$, the initial energy given is by $E_{i} = 178.2\;GeV-\Delta \tilde{E}(0,\omega_{c})$, and the final energy is given by $E_{f} = 178.2\;GeV- \Delta \tilde{E}(0,150)$. This parameterization ensures that all energy radiated below the cutoff frequency does not get included in the analysis. Finally, we assume the emission is time independent when evaluating the time integral of the power spectrum to be used in the energy emission $\Delta \tilde{E}(\omega_{c},\omega_{f})$, i.e. to integrate over time, we simply multiply the power spectrum by the total crystal crossing time. This is based on the time independence of the acceleration. However, we must take into account the thermalization time of the highest frequency emitted. From Fig. 1(a), the thermalization time of highest frequency photon emitted, $150$ GeV, is approximately $\tau_{150} = 8.45\times10^{-12} \;s$. Then, considering the $3.8$ mm thickness of the crystal, the total integration time is given by $\int dt = (3.8 \;mm)/c - \tau_{150}$. We must also note that although the area-entropy ratio converges to the appropriate value of $4 \ell_{p}^{2}$, there is still a very slight systematic slope which causes the ratio to diverge. This is most likely due to an additional hard photon emission process which is occurring that is not described by our power spectrum, i.e. it has a sufficiently low emission rate that prohibits it from thermalizing. This is also reflected in the oscillation of the chi-squared at the higher frequencies. 

In order to compute the Rindler bath temperature, we utilized the best fit parameters of the original theoretical power spectrum for each cutoff and applied them to the Rindler temperature-dependent power spectrum. We then performed best fits for each temperature at cutoffs from $30$ GeV to $100$ Gev in $10$ GeV steps, and then computed the resulting chi-squares to ensure that each were still within the 1 standard deviation threshold. These best fit temperatures were then averaged to produce a Rindler bath temperature of $T_{r} = 1.96 \pm .49 \; PeV$. When compared to the average FDU temperature $T_{FDU} = 1.80 \pm .51 \; PeV$, we find the measured temperature of Rindler bath to be $T_{r} = T_{FDU}(1.09 \pm .41)$. The error comes from the standard deviation from the mean.

\subsection{The AQED Response Function}

To begin our analysis we must first define the AQED response function. As we such, we examine electromagnetic emission in a refractive medium using the current interaction for QED

\bqe
\hat{S}_{I} = \int d^{4}x \hat{j}_{\m}(x)\hat{A}^{\m}(x).
\eqe

We shall couple an Unruh-DeWitt detector to the vector current. This will endow the electron an extra degree of freedom for energy transitions, i.e. the recoil. As such,

\bqe
\hat{j}_{\m}(x) = u_{\m}\hat{q}(\tau)\delta^{3}(x-x_{tr}(\tau)). 
\eqe

The monopole moment operator $\hat{q}(t)$ is Heisenberg evolved via $\hat{q}(\tau) = e^{i\hat{H}\tau}\hat{q}(0)e^{-i\hat{H} \tau}$ with $\hat{q}(0)$ defined as $\hat{q}(0)\ket{E_{i}} = \ket{E_{f}}$ with $E_{i}$ and $E_{f}$ the initial energy and final energy of a two level system moving along the trajectory, $x_{tr}(t)$, of the current; transitions both up and down energy are allowed. With the intent to examine Larmor radiation, both in vacuum and an optical medium, we formulate the following amplitude;

\bqe
\mathcal{A} = i\bra{\mathbf{k}}\otimes \bra{E_{f}}\hat{S}_{i}\ket{E_{i}}\otimes \ket{0}.
\eqe

The differential probability per unit final state momenta is given by, $\frac{d\mathcal{P}}{d^{3}k} = \vert \mathcal{A} \vert^{2}$. Evaluation yields

\bqa
\frac{d\mathcal{P}}{d^{3}k} &=& \vert  \bra{\mathbf{k}}\otimes \bra{E_{f}}\int d^{4}x \hat{j}_{\m}(x)\hat{A}^{\m}(x)\ket{E_{i}}\otimes \ket{0} \vert^{2} \non \\
&=& \int d^{4}x\int d^{4}x'\vert   \bra{E_{f}} \hat{j}_{\m}(x)\ket{E_{i}} \vert^{2} \vert \bra{\mathbf{k}} \hat{A}^{\m}(x)\ket{0} \vert^{2}.
\eqa

Note, the probability factorizes into an electron matrix element contracted with the photon matrix element. The electron matrix element yields

\bqa
\vert   \bra{E_{f}} \hat{j}_{\m}(x)\ket{E_{i}} \vert^{2} &=& \vert   \bra{E_{f}} u_{\m}(x)e^{i\hat{H}\tau}\hat{q}(0)e^{-i\hat{H} \tau}\delta^{3}(x-x(\tau))\ket{E_{i}} \vert^{2} \non \\
&=& q^{2} U_{\m \nu}[x',x]\delta^{3}(x-x_{tr}(\tau))\delta^{3}(x'-x'_{tr}(\tau')) e^{-i\Delta E(\tau'-\tau)} 
\label{current}
\eqa

Here we have defined the energy gap as $\Delta E = E_{f} -E_{i}$ and the charge as $q^{2} =\vert \bra{E_{f}} \hat{q}(0)\ket{E_{i}} \vert^{2}$. For the sake of brevity we defined a ``velocity tensor" via $U_{\m \nu}[x',x] = u_{\nu}(x')u_{\m}(x)  $. Next, we shall evaluate the photon inner product. For this we will need to integrate over the final state momenta, thereby developing the total emission probability. Hence,

\bqa
\int d^{3}k  \vert \bra{\mathbf{k}} \hat{A}^{\m}(x)\ket{0} \vert^{2} &=& \int d^{3}k \bra{0} \hat{A}^{\dagger \nu}(x')\ket{\mathbf{k}}\bra{\mathbf{k}} \hat{A}^{\m}(x)\ket{0} \non \\
&=& \bra{0} \hat{A}^{\dagger \nu}(x')\hat{A}^{\m}(x)\ket{0} \non \\
&=& G^{\nu \m}[x',x].
\eqa

Note we have utilized the completeness relation, $\int dk \ket{k}\bra{k} = 1$, to simplify the expression . The resultant is our photon two point function with vector indices. Using our photon two point function and the electron current density from Eq. (\ref{current}). we can formulate the AQED response function $\frac{d\mathcal{P}}{d \eta} = \Gamma$. Hence

\bqa
 \mathcal{P} &=& \int d^{3}k\int d^{4}x\int d^{4}x'\vert   \bra{E_{f}} \hat{j}_{\m}(x)\ket{E_{i}} \vert^{2} \vert \bra{\mathbf{k}} \hat{A}^{\m}(x)\ket{0} \vert^{2} \non\\
\Rightarrow \Gamma &=& q^{2} \int d\xi e^{-i\Delta E \xi}  U_{\m \nu}[x',x]G^{\nu \m}[x',x].
\eqa

Here we have made use of the difference and average propertime change of variables; $\xi = \tau' - \tau$ and $\eta = (\tau' + \tau)/2$ respectively. Using the standard mode decomposition for the vector field in a dielectric medium, we have

\bqa
\hat{A}^{\m}(x) &=& \int \frac{d^{3}k}{(2 \pi)^{3/2}} \frac{\sum_{i}\epsilon_{i}^{\m}}{\sqrt{2\omega}} \lbk \hat{a}_{k}e^{i(\mathbf{k}\cdot \mathbf{x} - \omega t) } + \hat{a}^{\dagger}_{k}e^{-i(\mathbf{k}\cdot \mathbf{x} - \omega t)}  \rbk \non \\
&=& \int \frac{d^{3}k}{(2 \pi)^{3/2}} \frac{\sigma^{\m}}{\sqrt{2\omega}} \lbk \hat{a}_{k}e^{i(\mathbf{k}\cdot \mathbf{x} - \omega t) } + \hat{a}^{\dagger}_{k}e^{-i(\mathbf{k}\cdot \mathbf{x} - \omega t)}  \rbk.
\eqa

In the last line we have defined the quantity $\sigma^{\m} = \sum_{i}\epsilon_{i}^{\m}$. The two point function then reduces to an integral over the momentum,

\bqa
\bra{0} \hat{A}^{\dagger \nu}(x')\hat{A}^{\m}(x)\ket{0} &=& \bra{0} \int \frac{d^{3}k'}{(2 \pi)^{3/2}} \frac{\sigma^{'\dagger \nu}}{\sqrt{2\omega'}}\lbk \hat{a}_{k'}e^{i(\mathbf{k'}\cdot \mathbf{x'} - \omega' t') } + \hat{a}^{\dagger}_{k'}e^{-i(\mathbf{k'}\cdot \mathbf{x'} - \omega' t')}  \rbk \non \\
&\;\;\;\;\; \times& \int \frac{d^{3}k}{(2 \pi)^{3/2}} \frac{\sigma^{\m}}{\sqrt{2\omega}} \lbk \hat{a}_{k}e^{i(\mathbf{k}\cdot \mathbf{x} - \omega t) } + \hat{a}^{\dagger}_{k}e^{-i(\mathbf{k}\cdot \mathbf{x} - \omega t)}  \rbk \ket{0} \non \\
&=& \frac{1}{(2 \pi)^{3}}\frac{1}{2} \int \frac{d^{3}k}{\omega}\sigma^{\m} \sigma^{\dagger \nu} e^{i(\mathbf{k}\cdot \Delta \mathbf{x} - \omega (t'-t))}.
\eqa

Again we see that the vector two point function is formally the same as a scalar field but with polarization vectors lending their indices. Combining all the pieces we can formulate the response function for our photon emission. 

\bqe
\Gamma = q^{2}  \frac{1}{(2 \pi)^{3}}\frac{1}{2}\int d\xi \int \frac{d^{3}k}{\omega} U e^{-i(\Delta E \xi - \mathbf{k}\cdot \Delta \mathbf{x}_{tr} + \omega \Delta t)}.
\eqe

We defined the velocity product $U = \sigma^{\m} \sigma^{\dagger \nu}  U_{\m \nu}[x',x]$ for brevity. Let us now examine the power radiated by a uniformly accelerated charge.

\subsection{The Thermalized Larmor Formula}

To analyze Larmor emission we will now consider the electron propagating in free space, i.e. $n=1$. We will begin by examining the polarization vectors that are contracted with our velocity tensor. Recalling that under proper acceleration $a$, the four-velocities at proper time $\tau$ will be given $u^{\m} = (\cosh{(a \tau)},0,0,\sinh{(a \tau)})$. Hence,

\bqa
U &=& \lb \sum_{i} u_{\m}\epsilon^{\m}_{i} \rb \lb \sum_{j} u_{\nu}\epsilon^{\nu}_{j} \rb^{\dagger} \non \\
&=& \lb \sum_{i} \mathbf{u}\cdot\mathbf{\epsilon}_{i} \rb \lb \sum_{j} \mathbf{u}'\cdot\mathbf{\epsilon}_{j}  \rb^{\dagger} \non \\
&=&\sinh{(a\tau)} \sinh{(a\tau')} \sin^{2}{(\theta)}.
\eqa

Here $\theta$ is the angle of photon emission relative to the direction of propagation along the $z$-axis. Moreover we will make use of the hyperbolic double angle formulas to obtain 

\bqe
\sinh{(a\tau)} \sinh{(a\tau')}  = \frac{1}{2}\lbk 2\cosh^{2}{(a\eta)} -1 - \cosh{(a\xi)} \rbk.
\eqe

Combining all the above pieces we can now formulate the transition probability. Thus

\bqa
\Gamma &=& q^{2}  \frac{1}{(2 \pi)^{3}}\frac{1}{4}\int d\xi \int \frac{d^{3}k}{\omega} \lbk 2\cosh^{2}{(a\eta)} -1 - \cosh{(a\xi)} \rbk  \sin^{2}{(\theta)} e^{-i(\Delta E \xi - \mathbf{k}\cdot \Delta \mathbf{x}_{tr} + \omega \Delta t)} .
\eqa

To arrive at the Larmor formula, typically computed in the rest frame of the electron, we follow \cite{jackson} and set $\beta = 0$; i.e. $\Delta x \ll \Delta t$. Note that with respect to the variable $\eta$, the Lorentz gamma is given by $\gamma = \cosh{(a\eta)}$ which we also take to be $1$. As such we obtain

\bqa
\Gamma &=& q^{2}  \frac{1}{(2 \pi)^{3}}\frac{1}{4}\int d\xi \lbk 1 - \cosh{(a\xi)} \rbk \int \frac{d^{3}k}{\omega}   \sin^{2}{(\theta)} e^{-i(\Delta E \xi  + \omega \Delta t)}.
\label{nrate}
\eqa

Now, examining the momentum integrations, we move to spherical coordinates with the momentum aligned along the z-axis, to yield

\bqa
\Gamma &=& q^{2}  \frac{1}{(2 \pi)^{3}}\frac{1}{4}\int d\xi \lbk 1 - \cosh{(a\xi)} \rbk \int d\theta d\omega d\phi \omega  \sin^{3}{(\theta)} e^{-i(\Delta E \xi  + \omega \Delta t)} \non \\
&=& \frac{2}{3}\alpha  \frac{1}{2 \pi}\int d\xi \lbk 1 - \cosh{(a\xi)} \rbk \int  d\omega  \omega    e^{-i(\Delta E \xi  + \omega \Delta t)} 
\eqa

Note, in the last line we rewrote the prefactor in terms of the fine structure constant $\alpha = \frac{q^{2}}{4 \pi}$.To compute the Larmor formula, we will need to examine the power emitted by the photon. As such we weight the frequency integral with an additional factor of frequency. Hence

\bqa
\mathcal{S} = \frac{2}{3}\alpha  \frac{1}{2 \pi}\int d\xi e^{-i\Delta E\xi} \lbk 1 - \cosh{(a\xi)} \rbk  \int  d\omega  \omega^{2}  e^{-i \omega \Delta t}. 
\eqa

The integration over the frequency can now be carried out to yield

\bqe
\int  d\omega  \omega^{2}   e^{-i \omega \Delta t} = \frac{2i}{\Delta t^{3}}.
\eqe

We should note that there is an implicit regulator $\Delta t \rightarrow \Delta t-i\epsilon$ in the denominator. This will later require us to include a pole on the real axis in the integration over the proper time. Our integration now simplifies to

\bqa
\mathcal{S} = \frac{2}{3}\alpha  \frac{i}{\pi}\int d\xi e^{-i\Delta E\xi} \frac{\lbk 1 - \cosh{(a\xi)} \rbk}{\Delta t^{3}}. 
\eqa

Finally we recall that $\Delta t = \frac{2}{a}\sinh{(a\xi/2)}$, we have

\bqa
\mathcal{S} = \frac{2}{3}\alpha  \frac{i}{\pi}\lb \frac{a}{2}\rb^{3}\int d\xi e^{-i\Delta E\xi} \frac{\lbk 1 - \cosh{(a\xi)} \rbk}{\sinh^{3}{(a\xi/2)}}. 
\eqa

Converting the hyperbolic terms to exponentials and making the change of variables $w = e^{a\xi}$, we have

\bqa
\mathcal{S} = \frac{2}{3}\alpha  \frac{i}{\pi}\lb \frac{a}{2}\rb^{3} \frac{8}{a} \int dw \frac{\lbk w^{1/2-i\Delta E/a} -\frac{1}{2}w^{3/2-i\Delta E/a} -\frac{1}{2} w^{-1/2-i\Delta E/a}   \rbk}{\lbk w-1   \rbk^{3}}. 
\eqa

This integration is standard and can be evaluated using the residue theorem. As such we obtain the following 

\bqe
\mathcal{S}= \frac{2}{3}\alpha a^{2} \frac{1}{1+e^{2\pi \Delta E/a}}.
\eqe

This is our thermal Larmor formula. By summing over transitions both up and down in energy, i.e. when $\Delta E = |\Delta E|$ and $\Delta E = -|\Delta E|$, and taking the limit $|\Delta E| \rightarrow 0$ we arrive at the standard Larmor formula. Hence

\bqe
\mathcal{S} = \frac{2}{3}\alpha a^{2}.
\eqe

Note, this is written in terms of the proper acceleration and therefore is fully relativistic as in the classical derivation.

\subsection{The Quantum Correction to the Larmor Formula}

The quantum correction to the Larmor formula will come from the recoil correction. Generically, by setting the energy gap to $\frac{\omega^{2}}{2m}$ we can examine what the quantum correction will be. Let us consider Eq. (S15) of the supplementary.

\bqa
\Gamma &=& \frac{2}{3}\alpha  \frac{1}{2 \pi}\int d\xi \lbk 1 - \cosh{(a\xi)} \rbk \int  d\omega  \omega    e^{-i(\Delta E \xi  + \omega \Delta t)} 
\eqa

Defining $\Delta \bar{E}$ as the auxiliary gap, the total energy gap we will use is defined as follows;

\bqe
\Delta E = \frac{\omega^2}{2m} + \Delta \bar{E}.
\eqe  

With this energy gap we note that the recoil portion must be integrated over with the frequency while the auxiliary gap serves the same role as a traditional Unruh-DeWitt detector. As such we split the detector up into the two portions; the frequency dependent part and the auxiliary gap. Thus,

\bqa
\Gamma &=& \frac{2}{3}\alpha  \frac{1}{2 \pi}\int d\xi \lbk 1 - \cosh{(a\xi)} \rbk  \int  d\omega  \omega   e^{-i(\Delta E \xi  + \omega \Delta t)} \non \\
&=& \frac{2}{3}\alpha  \frac{1}{2 \pi}\int d\xi e^{-i\Delta \bar{E}\xi}\lbk 1 - \cosh{(a\xi)} \rbk \int  d\omega  \omega    e^{-i(\frac{\omega^2}{2m}\xi  + \omega \Delta t)}. 
\eqa

To compute the Larmor formula, we will need to examine the power emitted by the photon. As such we weight the frequency integral with an additional factor of frequency. Hence

\bqa
\mathcal{S} = \frac{2}{3}\alpha  \frac{1}{2 \pi}\int d\xi e^{-i\Delta \bar{E}\xi} \lbk 1 - \cosh{(a\xi)} \rbk  \int  d\omega  \omega^{2}  e^{-i(\frac{\omega^2}{2m}\Delta t  + \omega \Delta t)}. 
\eqa

Note that since we are in a non relativistic regime we can make the replacement $\xi \rightarrow \Delta t$ in the frequency integration. Moreover, to evaluate the resultant integrals we Taylor expand the exponent that is quadratic in the frequency to yield

\bqe
\mathcal{S} = \frac{2}{3}\alpha  \frac{1}{2 \pi}\int d\xi e^{-i\Delta \bar{E}\xi}\lbk 1 - \cosh{(a\xi)} \rbk \sum_{\ell = 0}^{\infty } \frac{ \lb \frac{-i\Delta t}{2m} \rb^{\ell}}{\ell !}   \int  d\omega  \omega^{2(\ell + 1)}   e^{-i \omega \Delta t}. 
\eqe

The integrations over the frequency can now be carried out using the identity

\bqe
\int  d\omega  \omega^{2(\ell + 1)}   e^{-i \omega \Delta t} = \frac{(-1)^{2\ell + 3}(2\ell+2)!}{(-i\Delta t)^{2\ell +3}}.
\eqe

We should note that there is an implicit regulator $\Delta t \rightarrow \Delta t-i\epsilon$ in the denominator. This will later require us to include a pole on the real axis in the integration over the proper time. Our integration now simplifies to

\bqa
\mathcal{S} &=& \frac{2}{3}\alpha  \frac{1}{2 \pi}\int d\xi e^{-i\Delta \bar{E}\xi}\lbk 1 - \cosh{(a\xi)} \rbk \sum_{\ell = 0}^{\infty } \frac{ \lb \frac{-i\Delta t}{2m} \rb^{\ell}}{\ell !}   \int  d\omega  \omega^{2(\ell + 1)}   e^{-i\omega \Delta t} \non \\
&=& \frac{2}{3}\alpha  \frac{1}{2 \pi} \sum_{\ell = 0}^{\infty } \frac{(2\ell +2)! }{(-2m)^{\ell}\ell !} \frac{1}{(i)^{\ell +3}}\int d\xi \frac{e^{-i\Delta \bar{E}\xi}\lbk 1 - \cosh{(a\xi)} \rbk}{(i\Delta T)^{\ell +3}}.    
\eqa

Finally we recall that $\Delta t = \frac{2}{a}\sinh{(a\xi/2)}$, we have

\bqa
\mathcal{S} &=& \frac{2}{3}\alpha  \frac{1}{2 \pi} \sum_{\ell = 0}^{\infty } \frac{(2\ell +2)! }{(-2m)^{\ell}\ell !}  \lb \frac{a}{2i} \rb^{\ell +3} \int d\xi \frac{e^{-i\Delta \bar{E}\xi}\lbk 1 - \cosh{(a\xi)} \rbk}{\sinh^{\ell +3}{(a\xi/2)}}.    
\eqa

These integrations are standard and can be evaluated on a case by case basis. For the case of $\ell = 0$ we obtain the following 

\bqe
\mathcal{S}_{0} = \frac{2}{3}\alpha a^{2} \frac{1}{1+e^{2\pi \Delta \bar{E}/a}}.
\eqe

By summing over transition up and down in energy, i.e. when $\Delta \bar{E} = |\Delta \bar{E}|$ and $\Delta \bar{E} = -|\Delta \bar{E}|$, and taking the limit $|\Delta\bar{E}| \rightarrow 0$ we arrive at the Larmor formula. Hence

\bqe
\mathcal{S}_{0} = \frac{2}{3}\alpha a^{2}.
\eqe

For the first quantum correction to Larmor we evaluate the $\ell = 1$ integral. As such we obtain

\bqe
\mathcal{S}_{1} = \frac{4\alpha}{m} a^{2} \frac{\Delta \bar{E}}{1-e^{2\pi \Delta \bar{E}/a}}.
\eqe

Then by summing over transition up and down in energy we obtain

\bqa
\mathcal{S}_{1} &=& \frac{4\alpha}{m} a^{2}\Delta \bar{E} \frac{\sinh{(2\pi \Delta \bar{E}/a)}}{1-\cosh{(2\pi \Delta \bar{E}/a)}} \non \\
&=& -\frac{4\alpha}{\pi m} a^{3} \non \\
&=& -\frac{8\alpha}{ m} a^{2}T_{FDU}.
\eqa

Note, in the last line we took the limit of zero auxiliary energy gap. This is the quantum correction to the Larmor formula. Next we compute the radiation reaction force for each term. 

\bqa
\int F^{rr} v\cdot dt &=& -\int \mathcal{S} dt \non \\
\Rightarrow F^{rr}_{0} &=& \frac{2}{3}\alpha J \non \\
\Rightarrow F^{rr}_{1} &=& -\frac{8\alpha}{\pi m} J a.
\eqa

As such we find the covariant form of the quantum LAD equation to be

\bqe
m\frac{du^{\m}}{ds} = qF^{\m\nu}u_{\nu} + \frac{2}{3}\alpha\hbar \lb 1-\frac{24\hbar}{\pi mc^{3}} \sqrt{a^{2}} \rb\lbk J^{\m}+a^2u^{\m}  \rbk.
\eqe

\subsection{Power Spectrum}

Prior to integrating over the emitted photons frequency, Eq. (\ref{nrate}), we will have the following emission rate  

\bqa
\Gamma &=& q^{2}  \frac{1}{(2 \pi)^{3}}\frac{1}{4}\int d\xi \int \frac{d^{3}k}{\omega} \lbk 2\cosh^{2}{(a\eta)} -1 - \cosh{(a\xi)} \rbk  \sin^{2}{(\theta)} e^{-i(\Delta E \xi - \mathbf{k}\cdot \Delta \mathbf{x}_{tr} + \omega \Delta t)}.\label{pow}
\eqa

We will further simplify by using the following redefinition $\delta = 2 \gamma^2 - 1$. Utilizing the same approximation $\Delta x \sim 0$, which reproduces the Larmor formula, but now keeping our boost parameter, which is a function of $\eta$ only, we can now convert to spherical coordinates and integrate over the emission angle. Hence

\bqa
\Gamma &=& q^{2}  \frac{1}{(2 \pi)^{3}}\frac{1}{4}\int d\xi \int \frac{d^{3}k}{\omega} \lbk \delta - \cosh{(a\xi)} \rbk  \sin^{2}{(\theta)} e^{-i(\Delta E \xi  + \omega \Delta t)}  \non \\
&=& q^{2}  \frac{1}{(2 \pi)^{2}}\frac{1}{3}\int d\xi \int d\omega  \omega \lbk \delta - \cosh{(a\xi)} \rbk e^{-i(\Delta E \xi + \omega \Delta t)}  \non \\
\eqa

weighting by an extra factor of frequency to obtain the power, we formulate the power spectrum, $\frac{d \mathcal{S}}{d \omega}$. Hence 

\bqa
\frac{d \mathcal{S}}{d \omega}&=& q^{2}  \frac{1}{(2 \pi)^{2}}\frac{1}{3}\int d\xi   \omega^2 \lbk \delta - \cosh{(a\xi)} \rbk e^{-i(\Delta E \xi + \omega \Delta t)} 
\eqa

Recalling that with the boost parameter, $\Delta t = \frac{2}{a}\sinh{(a\xi/2)} \gamma$, we now convert hyperbolic cosine to exponentials to obtain

\bqa
\frac{d \mathcal{S}}{d \omega}= q^{2} \frac{1}{(2 \pi)^{2}}\frac{ \omega^2}{3}\int d\xi  &\Bigg[&  \delta e^{-i(\Delta E \xi + \frac{2\omega \gamma}{a}\sinh{(a\xi/2)} )} \non \\
&-&\frac{1}{2} e^{-i( (\Delta E +ia) \xi + \frac{2\omega \gamma}{a}\sinh{(a\xi/2)} )} \non \\
&-&\frac{1}{2} e^{-i( (\Delta E -ia) \xi + \frac{2\omega \gamma}{a}\sinh{(a\xi/2)} )} \Bigg]. 
\eqa

Employing the change of variables $w = a\xi/2$ we obtain

\bqa
\frac{d \mathcal{S}}{d \omega}= q^{2} \frac{1}{(2 \pi)^{2}}\frac{ \omega^2}{3} \frac{2}{a}\int dw  &\Bigg[&  \delta e^{(-\frac{2i\Delta E}{a}w - \frac{2i\omega \gamma}{a}\sinh{(w)} )} \non \\
&-&\frac{1}{2} e^{( -(\frac{2i\Delta E}{a} -2) w - \frac{2i\omega \gamma}{a}\sinh{(w)} )} \non \\
&-&\frac{1}{2} e^{( -(\frac{2i\Delta E}{a} +2) w - \frac{2i\omega \gamma}{a}\sinh{(w)} )} \Bigg]. 
\eqa

Now, recalling the integral representation of the second Hankel function, we have

\bqa
H^{(2)}_{n}(x) = -\frac{1}{i\pi}\int_{-\infty}^{\infty -i \pi} dt e^{-nt+x\sinh(t)}.\label{hank}
\eqa

Here, the integration contour is shifted down by $-\pi$ on the imaginary axis. This is consistent with the Larmor case since there we used our $\Delta t - i \epsilon$ prescription. Using the above formula we find our power spectrum to be

\bqa
\frac{d \mathcal{S}}{d \omega}= -i \frac{2}{3}\alpha \frac{ \omega^2}{a}  \lbk  \delta H^{(2)}_{\frac{2i\Delta E}{a}}\lb - \frac{2i\omega \gamma}{a} \rb  -\frac{1}{2} \lb  H^{(2)}_{\frac{2i\Delta E}{a} -2}\lb - \frac{2i\omega \gamma}{a} \rb +  H^{(2)}_{\frac{2i\Delta E}{a} +2}\lb - \frac{2i\omega \gamma}{a} \rb \rb \rbk. \label{power1}
\eqa

To make the presence of thermality more apparent, we make use of the following identity, $H^{(2)}_{\ell}(x) = e^{i\ell \pi}H^{(2)}_{-\ell}(x)$. As such, each term will yield precisely a Boltzmann factor with the Unruh-DeWitt detector energy gap thermalized at the celebrated Fulling-Davies-Unruh temperature. Applying this identity then yields our power spectrum,

\bqa
\frac{d \mathcal{S}}{d \omega}= -i \frac{2}{3}\alpha \frac{ \omega^2}{a}  e^{-2\pi \Delta E/a}  \lbk  \delta H^{(2)}_{-\frac{2i\Delta E}{a}}\lb - \frac{2i\omega \gamma}{a} \rb  -\frac{1}{2} \lb  H^{(2)}_{-\frac{2i\Delta E}{a} +2}\lb - \frac{2i\omega \gamma}{a} \rb +  H^{(2)}_{\frac{-2i\Delta E}{a} -2}\lb - \frac{2i\omega \gamma}{a} \rb \rb \rbk. \label{power2}
\eqa

This power spectrum contains the Unruh-DeWitt detector energy gap thermalized at the Fulling-Davies-Unruh temperature. Let us now explore the use of radiative energy loss, as a source for acceleration. We will then compare our results with the recent experimental observation of radiation reaction in aligned crystals.

\subsection{The Thermalized Total Transverse Emission Spectrum}

In order to understand the physical processes present in the Rindler frame we must first start by computing the emission spectrum per unit transverse momentum, i.e. $\frac{d\Gamma}{dk_{\perp}}$. In the Rindler frame, the mode functions of the photons are labeled by momentum transverse to the acceleration and by the Rindler frequency $\omega_{r}$ which characterizes the photons energy in Rindler frame. We will find that $\omega_{r}$ will play the role of the energy gap of the Unruh-DeWitt detector. Moreover, there is no dispersion relation which relates $\omega_{r}$ to $k_{\perp}$. For this reason, we must utilize our formalism to compute $\frac{d\Gamma}{d^{2}k_{\perp}}$ in both the Rindler frame and the lab frame to confirm that they match. Starting with the lab frame computation, let us recall Eq. (\ref{pow}),

\bqa
\Gamma &=& q^{2}  \frac{1}{(2 \pi)^{3}}\frac{1}{4}\int d\xi \int \frac{d^{3}k}{\omega} \lbk 2\cosh^{2}{(a\eta)} -1 - \cosh{(a\xi)} \rbk  \sin^{2}{(\theta)} e^{-i(\Delta E \xi - \mathbf{k}\cdot \Delta \mathbf{x}_{tr} + \omega \Delta t)} \non \\
\frac{d\Gamma}{d^{2}k_{\perp}} &=& q^{2}  \frac{1}{(2 \pi)^{3}}\frac{1}{4}\int d\xi \int \frac{dk_{z}}{\omega} \lbk\delta - \cosh{(a\xi)} \rbk  \sin^{2}{(\theta)} e^{-i(\Delta E \xi - \mathbf{k}\cdot \Delta \mathbf{x}_{tr} + \omega \Delta t)}
\eqa 

Note again we made use of the boost factor $\delta = 2\cosh^{2}{(a\eta)} -1$ and we have defined $d^{2}k_{\perp} = dk_{x}dk_{y}$. Using the same approximation $\Delta x \sim 0$ and recalling $\Delta t = \frac{2}{a}\sinh{(a\xi/2)} \gamma$, we have

\bqa
\frac{d\Gamma}{d^{2}k_{\perp}} &=& q^{2}  \frac{1}{(2 \pi)^{3}}\frac{1}{4}\int d\xi \int \frac{dk_{z}}{\omega} \lbk\delta - \cosh{(a\xi)} \rbk  \sin^{2}{(\theta)} e^{-i(\Delta E \xi + \frac{2\omega \gamma}{a}\sinh{(a\xi/2)})}.
\eqa 

Utilizing the same Hankel identity, Eq. (\ref{hank}), and recalling $\sin{(\theta)} = \frac{k_{\perp}}{\omega}$ we arrive at,

\bqa
\frac{d\Gamma}{d^{2}k_{\perp}} &=& \frac{-i\alpha}{4\pi a} \int dk_{z}\frac{k_{\perp}^{2}}{\omega^{3}} \Big[ \delta H^{(2)}_{\frac{2i\Delta E}{a}}\lb - \frac{2i\omega \gamma}{a} \rb \\ \non
 & -& \frac{1}{2} \lb  H^{(2)}_{\frac{2i\Delta E}{a} -2}\lb - \frac{2i\omega \gamma}{a} \rb +  H^{(2)}_{\frac{2i\Delta E}{a} +2}\lb - \frac{2i\omega \gamma}{a} \rb \rb \Big].
\eqa 

Finally, if we recall that $\omega = \sqrt{k_{\perp}^{2}+k_{z}^{2}}$ and assume that the functional dependence on $k_{z}$ in the argument of the Hankel functions is negligible, i.e. $\omega \sim k_{\perp}$, then the integral over $k_{z}$ can be evaluated. Since each photon from a thermalized process will be emitted into Rindler horizon. This occurs due to diphoton creation near the Rindler horizon which causes one to escape into the horizon, while the other is absorbed by the electron or by direct photon emission into the horizon; i.e. emission and absorption of Rindler photons \cite{matsas5,kolekar}. This implies the momentum of all thermalized photons, in this setting, will have $k_{z}\geq 0$. This fact may be of pertinence to understanding the nature of the relationship between the Minkowski photon energy $\omega$ and the Rindler photon frequency $\omega_{r}$. Hence

\bqa
\int dk_{z}\frac{k_{\perp}^{2}}{\omega^{3}} &=& \int_{0}^{\infty} dk_{z}\frac{k_{\perp}^{2}}{\lb k_{\perp}^{2} +k_{z}^{2} \rb^{3/2}} \non \\
 &=& 1.
\eqa

From here, we obtain our emission rate per unit transverse momentum,

\bqa
\frac{d\Gamma}{d^{2}k_{\perp}} &=& \frac{-i\alpha}{4\pi a} \Big[ \delta H^{(2)}_{\frac{2i\Delta E}{a}}\lb - \frac{2i k_{\perp} \gamma}{a} \rb \\ \non
 &-& \frac{1}{2} \lb  H^{(2)}_{\frac{2i\Delta E}{a} -2}\lb - \frac{2ik_{\perp} \gamma}{a} \rb +  H^{(2)}_{\frac{2i\Delta E}{a} +2}\lb - \frac{2ik_{\perp} \gamma}{a} \rb \rb \Big].
\eqa 

Finally, we recall that each Hankel function, by flipping the sign in the index, produces the detailed balance relationship. From this, when we can form the total emission rate, $\frac{d\Gamma_{tot}}{d^{2}k_{\perp}} = \frac{d\Gamma_{\Delta E}}{d^{2}k_{\perp}}+\frac{d\Gamma_{-\Delta E}}{d^{2}k_{\perp}}$. This yields,

\bqa
\frac{d\Gamma_{tot}}{d^{2}k_{\perp}} &=& \frac{d\Gamma_{\Delta E}}{d^{2}k_{\perp}}+\frac{d\Gamma_{-\Delta E}}{d^{2}k_{\perp}} \non \\
&=& \frac{d\Gamma_{\Delta E}}{d^{2}k_{\perp}}+e^{2\pi \Delta E/a}\frac{d\Gamma_{\Delta E}}{d^{2}k_{\perp}} \non \\
&=& \frac{d\Gamma}{d^{2}k_{\perp}}\lbk 1+ e^{2\pi \Delta E/a} \rbk.
\eqa 

Here, $\frac{d\Gamma}{d^{2}k_{\perp}}$, is given by Eq. (S47). Writing it explicitely for later comparion we find,

\bqa
\frac{d\Gamma_{tot}}{d^{2}k_{\perp}}&=& \frac{-i\alpha}{4\pi a}\lbk  \delta H^{(2)}_{\frac{2i\Delta E}{a}}\lb - \frac{2i k_{\perp} \gamma}{a} \rb  -\frac{1}{2} \lb  H^{(2)}_{\frac{2i\Delta E}{a} -2}\lb - \frac{2ik_{\perp} \gamma}{a} \rb +  H^{(2)}_{\frac{2i\Delta E}{a} +2}\lb - \frac{2ik_{\perp} \gamma}{a} \rb \rb \rbk \non \\
&\times & \lbk 1+ e^{2\pi \Delta E/a} \rbk. \label{mink}
\eqa

Note, that this also assumes that the energy gap, if dependent on the photon frequency $\omega$, also reduces to $\Delta E (\omega) \rightarrow \Delta E (k_{\perp})$. We must now ensure that the channeling radiation analysis, in the Rindler frame, matches the above expression.

\subsection{The Rindler Frame Analysis of a ``Channeling like" Oscillation}

The Rindler frame analysis provides us with a special opportunity to explore the particle content of the background spacetime and potential temperature experienced by an accelerated particle. In order to approach this computation, we must first look at our Rindler line element for the coordinates $(\tau,\xi,x,y)$; $ds^{2} = e^{2a\xi} \lb d\tau^{2}  - d\xi^{2} \rb - dx_{\perp}^{2}$. Here, $\xi$ characterizes motion along the z direction, $\tau$ is the Rindler time, and $x_{\perp}^{2}=x^{2}+y^{2}$. The two Rindler coordinates $(\tau,\xi)$ are related to the laboratory time and z cooridinate via; $t = (e^{a\xi}/a)\sinh{(a\tau)}$ and $z = (e^{a\xi}/a)\cosh{(a\tau)}$. Quantization of the electromagnetic field in the Rindler wedge yields the two physical photon modes,
\bqa
A^{1}_{\m}(x) &=& \frac{1}{2\pi^{2} k_{\perp}} \lb \frac{\sinh{(\pi \omega_{r}/a)}}{a} \rb^{1/2}\lb 0,0,k_{y}f(x),-k_{x}f(x)  \rb \non \\
A^{2}_{\m}(x) &=& \frac{1}{2\pi^{2} k_{\perp}} \lb \frac{\sinh{(\pi \omega_{r}/a)}}{a} \rb^{1/2}\lb \partial_{\xi}f(x),-i \omega_{r} f(x),0,0  \rb. 
\eqa
The transverse photon momentum $k_{x}$ and $k_{y}$ along with the mutually independent Rindler frequency $\omega_{r}$ label the Rindler photon mode functions; here $k_{\perp} = \sqrt{k_{x}^{2}+k_{y}^{2}}$. The function $f(x)$ which characterizes the spatial modulation of the modes is given by,
\bqe
f(x) = K_{i\omega_{r}/a}{\lb \frac{k_{\perp}}{a}e^{a\xi} \rb}e^{i(k_{\perp}\cdot x_{\perp}-\omega_{r}\tau)}.
\eqe
Our second quantized field operators are constructed from the above physics polarization modes, integrated over the transverse momentum and Rindler frequency,
\bqe
\hat{A}_{\m}(x) = \int_{\infty}^{\infty} d^{2}k_{\perp} \int_{0}^{\infty} d\omega_{r} \sum_{i=1,2}\lbk \hat{a}A^{i}_{\m}(x) + \hat{a}^{\dagger}A^{i\dagger}_{\m}(x)  \rbk.
\eqe
Here we are summing over both physical polarization modes. Now, in order to analyze any photon emission or absorption processes, we will use the curved spacetime QED interaction Lagrangian, $\mathcal{L} = \sqrt{-g}j^{\m}_{r}\hat{A}_{m}$. Here, the Rindler current will characterize the comoving channeling oscillation. The resulting amplitude $\mathcal{A}$ for Rindler photon absorption is given by
\bqe
\mathcal{A}_{abs} = i\int d^{4}x \sqrt{-g}j^{\m}_{r}\bra{0}\hat{A}_{\m}\ket{\gamma}.\label{amp}
\eqe
The magnitude for Rindler photon emission is the same for absorption, i.e. $|\mathcal{A}_{abs}|=|\mathcal{A}_{emi}|$. Moreover, if we consider the presence of a thermal backround in Rindler space, when we compute the total probability for both photon emission and absorption, we must take into account the background thermal bath. What this means is that when we compute the total probability, we must weight each component by the contribution from the thermal background, i.e. $\mathcal{P}_{abs} \sim |\mathcal{A}_{abs}|^{2}(1/e^{\omega_{r}/T}-1)$ and $\mathcal{P}_{emi} \sim |\mathcal{A}_{abs}|^{2}(1+1/(e^{\omega_{r}/T}-1))$. The temperature of the background thermal bath, $T$, is kept arbitrary. Summing both probabilities and integrating over the final Rindler photon states then gives the total probability,
\bqe
\mathcal{P} = \sum_{i=1,2}\int_{\infty}^{\infty} d^{2}k_{\perp} \int_{0}^{\infty} d\omega_{r}|\mathcal{A}^{i}_{abs}|^{2}\coth{(\omega_{r}/(2T))}.
\eqe
Now, in order to evaluate our absorption amplitude, we need the functional form of the Rindler current. The four velocity for a channeling like oscillation, e.g. a transverse oscillation in the x direction, can be written as,
\bqe
u^{\m} = \gamma_{x}(1,0,v_{0}\cos{(\Omega \tau)},0).
\eqe
Here the velocity parameter, $v_{0} = A \Omega$, is determined the oscillation amplitude, $A$, and oscillation frequency, $\Omega$. The Lorentz factor is determined from the transverse oscillation which we take to be non relativistic, i.e. $\gamma_{x} = 1$. From here we can construct our four current for the channeling oscillation is then given by
\bqe
j_{r}^{\m} = q u^{\m}\delta{(\xi)}\delta{(x-x_{tr})}\delta{(y)}.
\eqe 
The trajectory along the x direction is given by $x=A\sin{(\Omega \tau)}$ and $q$ is the electron charge. Utilizing this trajectory, we can examine the total rate of Rindler photon emission and absorption under the assumption that there is a background thermal bath present in the Rindler frame. Note, for the transverse oscillation, we will see that it is sufficient to only include the minimal oscillation term in the phase and keep the velocity term in the current constant, i.e. $v_{x} = v_{0}\cos{(\Omega \tau)} \sim v_{0}$ and $x_{tr} = A \sin{(\Omega \tau)} \sim v_{0}\tau$. We also must recall the covariant volume element is given by $\sqrt{-g} = e^{2a\xi}$. Recalling Eq. (\ref{amp}), let us compute the absorption amplitude for the first polarization. Hence,
\bqa
\mathcal{A}^{1}_{abs} &=& i\int d^{4}x \sqrt{-g}j^{\m}_{r}\bra{0}\hat{A}_{\m}^{1}\ket{\gamma} \non \\
&=& i\int d^{4}x \sqrt{-g}j^{\m}_{r}\int d^{2}k_{\perp} \int d\omega\bra{0}\lbk \hat{a}A^{1}_{\m}(x) + \hat{a}^{\dagger}A^{1\dagger}_{\m}(x)  \rbk\ket{\gamma} \non \\
&=& i\int d^{4}x  \sqrt{-g}j^{\m}_{r} A^{1}_{\m}(x)   \non \\
&=& i\int d^{4}x e^{2a\xi}\delta{(\xi)}\delta{(x-x_{tr})}\delta{(y)}u^{\m} A^{1}_{\m}(x)   \non \\
&=& \frac{-iq}{2\pi^{2}}\int d^{4}x e^{2a\xi}\delta{(\xi)}\delta{(x-x_{tr})}\delta{(y)} \frac{v_{0}k_{y}}{ k_{\perp}} \lb \frac{\sinh{(\pi \omega_{r}/a)}}{a} \rb^{1/2}K_{i\omega_{r}/a}{\lb \frac{k_{\perp}}{a}e^{a\xi} \rb}e^{i(k_{\perp}\cdot x_{\perp}-\omega_{r}\tau)}   \non \\
&=& \frac{-iq}{2\pi^{2}}\int d\tau \frac{v_{0}k_{y}}{ k_{\perp}} \lb \frac{\sinh{(\pi \omega_{r}/a)}}{a} \rb^{1/2}K_{i\omega_{r}/a}{\lb \frac{k_{\perp}}{a}\rb}e^{i(k_{x}v_{0} -\omega_{r})\tau}   \non \\
&=& \frac{-iq}{\pi} \frac{v_{0}k_{y}}{ k_{\perp}} \lb \frac{\sinh{(\pi \omega_{r}/a)}}{a} \rb^{1/2}K_{i\omega_{r}/a}{\lb \frac{k_{\perp}}{a}\rb}\delta{(\omega_{r}-k_{x}v_{0} )}. 
\eqa   
Similarly, for the second physical mode we obtain,
\bqa
\mathcal{A}^{2}_{abs} &=& i\int d^{4}x  \sqrt{-g}j^{\m}_{r} A^{2}_{\m}(x)   \non \\
&=& \frac{iq}{\pi}\lb \frac{\sinh{(\pi \omega_{r}/a)}}{a} \rb^{1/2}K^{'}_{i\omega_{r}/a}{\lb \frac{k_{\perp}}{a}\rb}\delta{(\omega_{r}-k_{x}v_{0} )}. 
\eqa   
Here, the derivative of the Bessel function is with respect to the argument. Now, by taking the magnitude squared of the above amplitudes, we can construct the total probability,
\bqa
\mathcal{P} &=& \int_{\infty}^{\infty} d^{2}k_{\perp} \int_{0}^{\infty} d\omega_{r} \lbk |\mathcal{A}^{1}_{abs}|^{2} + |\mathcal{A}^{2}_{abs}|^{2}   \rbk \coth{(\omega_{r}/(2T))} \non \\
&=& \frac{q^{2}}{\pi^{2}a}\int_{\infty}^{\infty} d^{2}k_{\perp} \int_{0}^{\infty} d\omega_{r}\sinh{(\pi \omega_{r}/a)}  \delta^{2}{(\omega_{r}-k_{x}v_{0} )}\coth{(\omega_{r}/(2T))} \non \\
&\times & \lbk  \lb \frac{v_{0}k_{y}}{ k_{\perp}} \rb^{2} |K_{i\omega_{r}/a}{\lb \frac{k_{\perp}}{a}\rb}|^{2} + |K^{'}_{i\omega_{r}/a}{\lb \frac{k_{\perp}}{a}\rb}|^{2} \rbk. \label{prob}
\eqa
We can now convert one of the delta functions into a total interaction time via $\Delta \tau = 2\pi\delta{(0)}$. This will allow us to formulate the emission rate, $\Gamma = \mathcal{P}/\Delta \tau$. Then, after integrating over the remaining delta function we fix the Rindler frequency to the channeling oscillation, $\omega_{r} = k_{x}A\Omega$. We can now formulate the emission rate per transverse momentum $\frac{d\Gamma}{d^{2}k_{\perp}} = $. Hence,
\bqa
\frac{d\Gamma}{d^{2}k_{\perp}} = \frac{q^{2}}{2\pi^{3}a}\sinh{(\pi \omega_{r}/a)}  \coth{(\omega_{r}/(2T))} \lbk  \lb \frac{v_{0}k_{y}}{ k_{\perp}} \rb^{2} |K_{i\omega_{r}/a}{\lb \frac{k_{\perp}}{a}\rb}|^{2} + |K^{'}_{i\omega_{r}/a}{\lb \frac{k_{\perp}}{a}\rb}|^{2} \rbk.
\eqa
We must note at this point that the integral over $k_{x}$ goes from $-\infty \rightarrow \infty$. This is because although we only kept the minimal contribution of the oscillation in the exponent, $sin{(\Omega \tau)} \rightarrow \Omega \tau$, the velocity still takes negative values in practice so we keep the full integration. Let us now consider the transformation of the K Bessel functions into Hankel functions. First, we must recall the identity $ K_{\alpha}(x)=\frac{\pi}{2}(-i)^{\alpha+1}H_{\alpha}^{(2)}{(-ix)} $. This identity allows us to write
\bqa
K_{i\omega_{r}/a}{\lb \frac{k_{\perp}}{a}\rb} & = & -i\frac{\pi}{2}e^{\frac{\pi \omega_{r}}{2a}}H^{(2)}_{i\omega_{r}/a}{\lb  -i\frac{k_{\perp}}{a}\rb} \non \\
K^{'}_{i\omega_{r}/a}{\lb \frac{k_{\perp}}{a}\rb} & = & -i\frac{\pi}{2}e^{\frac{\pi \omega_{r}}{2a}}H^{(2)'}_{i\omega_{r}/a}{\lb  -i\frac{k_{\perp}}{a}\rb}.
\eqa
From here, our emission spectrum transforms into,
\bqa
\frac{d\Gamma}{d^{2}k_{\perp}} &=& \frac{q^{2}}{8\pi a}\sinh{(\pi \omega_{r}/a)}  \coth{(\omega_{r}/(2T))} e^{\frac{\pi \omega_{r}}{a}} \non \\
& \times & \lbk  \lb \frac{v_{0}k_{y}}{ k_{\perp}} \rb^{2} |H^{(2)}_{i\omega_{r}/a}{\lb  -i\frac{k_{\perp}}{a}\rb}|^{2} + |H^{(2)'}_{i\omega_{r}/a}{\lb  -i\frac{k_{\perp}}{a}\rb}|^{2} \rbk.
\eqa
We can further reduce the above Hankel functions by recalling their integral identity, Eq. (\ref{hank}). Let us consider the first Hankel term in the above expression
\bqa
|H^{(2)}_{i\alpha}{\lb  -ix\rb}|^{2} &=& \frac{1}{-i \pi}\int dt e^{i\alpha t -ix\sinh{(t)}}\frac{1}{i \pi}\int dt' e^{-i\alpha t' + ix\sinh{(t')}} \non \\
&=& \frac{1}{ \pi^{2}}\int dt \int dt' e^{i\alpha t -ix\sinh{(t)} - i\alpha t' + ix\sinh{(t')}} \non \\
&=& \frac{1}{ \pi^{2}}\int dt \int dt' e^{-i\alpha(t'-t)  - ix(\sinh{(t')}-\sinh{(t)})} \non \\
&=& \frac{2}{ \pi^{2}}\int d\eta \int d\xi e^{-2i\alpha \xi   - 2ix\sinh{(\xi)}\cosh{(\eta)}} \non \\
&=& \frac{-2i}{ \pi}\int d\eta \int H^{(2)}_{2i\alpha}{\lb  -i2x\gamma\rb}.
\eqa
Here we made use of the change of variables $\xi = (t'-t)/2$ and $\eta = (t'+t)/2$. Moreover, note that we have recovered a Lorentz boost factor, $\gamma = \cosh{(\eta)}$ based on the rapidity variable $\eta$. Let us now evaluate the derivative term. To begin, we will transform the derivative back into pure Hankel functions via the use of the identity $H^{(2)'}_{i\alpha}{\lb  -ix\rb} = -\frac{i}{2}\lbk H^{(2)}_{i\alpha-1}{\lb  -ix\rb} -H^{(2)}_{i\alpha+1}{\lb  -ix\rb}  \rbk$. This will yield the intermediate step,
\bqa
|H^{(2)'}_{i\alpha}{\lb  -ix\rb}|^{2}&=&  \frac{1}{4} [ |H^{(2)}_{i\alpha-1}{\lb  -ix\rb}|^{2} + |H^{(2)}_{i\alpha+1}{\lb  -ix\rb}|^{2}  \non \\
& - & H^{(2)}_{i\alpha-1}{\lb  -ix\rb}H^{(2)\ast}_{i\alpha+1}{\lb  -ix\rb}-H^{(2)}_{i\alpha+1}{\lb  -ix\rb}H^{(2)\ast}_{i\alpha-1}{\lb  -ix\rb} ]
\eqa
Lets begin by evaluating the negative terms first. The first one reduces to,
\bqa
H^{(2)}_{i\alpha +1}{\lb  -ix\rb}H^{(2)\ast}_{i\alpha -1}{\lb  -ix\rb} & = & \frac{1}{-i \pi}\int dt' e^{-(i\alpha-1) t' -ix\sinh{(t')}}\frac{1}{i \pi}\int dt e^{-(-i\alpha +1) t + ix\sinh{(t)}} \non \\
& = & \frac{1}{\pi^{2}}\int dt \int dt' e^{-(i\alpha-1) t' -ix\sinh{(t')}-(-i\alpha +1) t + ix\sinh{(t)}} \non \\
& = & \frac{1}{\pi^{2}}\int dt \int dt' e^{-i\alpha t'- t' -ix\sinh{(t')}+i\alpha t+ t + ix\sinh{(t)}} \non \\
& = & \frac{2}{\pi^{2}}\int d \eta \int d \xi e^{-2(i\alpha +1)\xi -2ix\sinh{(\xi)}\cosh{(\eta)}} \non \\
& = & \frac{-2i}{ \pi}\int d\eta  H^{(2)}_{2i\alpha +2}{\lb  -i2x\gamma\rb}.
\eqa
Similary, evaluation of the other negative term yields,
\bqa
H^{(2)}_{i\alpha-1}{\lb  -ix\rb}H^{(2)\ast}_{i\alpha+1}{\lb  -ix\rb} & = & \frac{1}{-i \pi}\int dt' e^{-(i\alpha-1) t' -ix\sinh{(t')}}\frac{1}{i \pi}\int dt e^{-(-i\alpha +1) t + ix\sinh{(t)}} \non \\
&=& \frac{1}{ \pi^{2} }\int dt' \int dt e^{-i\alpha t' + t' -ix\sinh{(t')} + i\alpha t - t + ix\sinh{(t)}} \non \\
& = & \frac{1}{ \pi^{2} }\int dt' \int dt e^{-i\alpha (t' - t) + t' - t -ix(\sinh{(t')}-\sinh{(t)})} \non \\ 
& = & \frac{1}{ \pi^{2} }\int d\eta \int d \xi e^{-(i\alpha - 1)\xi -2ix\sinh{(\xi)}\cosh({\eta})} \non \\
& = & \frac{-2i}{ \pi}\int d\eta  H^{(2)}_{2i\alpha -2}{\lb  -i2x\gamma\rb}.
\eqa
Finally, we have only the remaining first two positive terms to evaluate. We will combine these two terms together when evaluating. Hence,
\bqa
 |H^{(2)}_{i\alpha-1}{\lb  -ix\rb}|^{2} + |H^{(2)}_{i\alpha+1}{\lb  -ix\rb}|^{2} &=& \frac{1}{-i \pi}\int dt e^{(i\alpha-1) t -ix\sinh{(t)}}\frac{1}{i \pi}\int dt' e^{(-i\alpha -1) t' + ix\sinh{(t')}} \non \\
 &+& \frac{1}{-i \pi}\int dt e^{(i\alpha+1) t -ix\sinh{(t)}}\frac{1}{i \pi}\int dt' e^{(-i\alpha +1) t' + ix\sinh{(t')}} \non \\
 &=& \frac{1}{ \pi^2}\int dt \int dt' e^{i\alpha t- t -ix\sinh{(t)} -i\alpha t' - t' + ix\sinh{(t')}} \non \\
 &+& \frac{1}{ \pi^2}\int dt \int dt' e^{i\alpha t+ t -ix\sinh{(t)} -i\alpha t' + t' + ix\sinh{(t')}} \non \\
&=& \frac{1}{ \pi^2}\int dt \int dt' e^{i\alpha t -ix\sinh{(t)} -i\alpha t'  + ix\sinh{(t')}} \lbk e^{-t-t'}+ e^{t+t'}  \rbk \non \\
&=& \frac{4}{ \pi^2}\int d\eta \int d \xi e^{-2i\alpha \xi -2ix\sinh{(\xi)}\cosh{(\eta)} } \cosh{(2\eta)} \non \\
&=& \frac{-4i}{  \pi}\int d\eta H^{(2)}_{2i\alpha}{\lb  -i2x\gamma\rb} \cosh{(2\eta)} \non \\
&=& \frac{-4i}{  \pi}\int d\eta \delta H^{(2)}_{2i\alpha}{\lb  -i2x\gamma\rb}.
\eqa
Note we have recovered the same boost factor as in the Minkowski frame case; $\delta = 2\gamma^2 - 1 = \cosh{(2\eta)}$. Let us now combine all the derivative terms together to yield,
\bqa
|H^{(2)'}_{i\alpha}{\lb  -ix\rb}|^{2} &=& \frac{1}{4}\int d\eta \lbk    \frac{-4i}{  \pi} \delta H^{(2)}_{2i\alpha}{\lb  -i2x\gamma\rb} - \frac{-2i}{ \pi}  H^{(2)}_{2i\alpha -2}{\lb  -i2x\gamma\rb} - \frac{-2i}{ \pi} H^{(2)}_{2i\alpha +2}{\lb  -i2x\gamma\rb} \rbk \non \\
&=&-\frac{i}{\pi}\int d\eta \lbk    \delta H^{(2)}_{2i\alpha}{\lb  -i2x\gamma \rb} - \frac{1}{2} \lb  H^{(2)}_{2i\alpha +2}{\lb  -i2x\gamma\rb} +H^{(2)}_{2i\alpha -2}{\lb  -i2x\gamma\rb}	\rb\rbk
\eqa
Now that we have all of our pieces together, we find our emission spectrum to be
\bqa
\frac{d\Gamma}{d^{2}k_{\perp}} &=& -\frac{i \alpha}{2\pi a}\int d\eta \sinh{(\pi \omega_{r}/a)}  \coth{(\omega_{r}/(2T))} e^{\frac{\pi \omega_{r}}{a}} \non \\
& \times & \Big[  \lb \delta + 2 \lb \frac{v_{0}k_{y}}{ k_{\perp}} \rb^{2} \rb H^{(2)}_{2i\omega_{r}/a}{\lb  \frac{-2ik_{\perp} \gamma }{a}\rb}\\ \non
&-& \frac{1}{2} \lb  H^{(2)}_{2i\omega_{r} +2}{\lb\frac{ -2ik_{\perp} \gamma }{a}\rb} +H^{(2)}_{2i\omega_{r} -2}{\lb \frac{-2ik_{\perp} \gamma }{a}\rb}	\rb \Big] .
\eqa
If we are considering a nonrelativistic channeling oscillation, then the term proportional to $\lb \frac{v_{0}k_{y}}{ k_{\perp}} \rb^{2}$ will be negligible. Moreover, note that we are integrating the above expression over the rapidity $\eta$. Restricting our emission spectrum to a window of constant rapidity, i.e. we maintain the assumption of constant acceleration and velocity as in the case applied to the experiment, we then obtain our total emission spectrum in the Rindler frame,
\bqa
\frac{d\Gamma}{d^{2}k_{\perp}} &=& -\frac{i \alpha}{2\pi a} \sinh{(\pi \omega_{r}/a)}  \coth{(\omega_{r}/(2T))} e^{\frac{\pi \omega_{r}}{a}} \non \\
& \times & \lbk    \delta  H^{(2)}_{2i\omega_{r}/a}{\lb  \frac{-2ik_{\perp} \gamma }{a}\rb}- \frac{1}{2} \lb  H^{(2)}_{2i\omega_{r} +2}{\lb\frac{ -2ik_{\perp} \gamma }{a}\rb} +H^{(2)}_{2i\omega_{r} -2}{\lb \frac{-2ik_{\perp} \gamma }{a}\rb}	\rb \rbk.
\eqa
In order to confirm that this expression matches the Minkowski expression, we must set the temperature of the background Rindler bath to the FDU temperature, $T = \frac{a}{2\pi}$, and compare. As such the temperature prefactor in the above expression reduces to,
\bqa
\sinh{(\pi \omega_{r}/a)}  \coth{(\omega_{r}/(2T))} e^{\frac{\pi \omega_{r}}{a}} &= &\sinh{(\pi \omega_{r}/a)}  \coth{(\pi\omega_{r}/(a))} e^{\frac{\pi \omega_{r}}{a}} \non \\
&= &\cosh{(\pi \omega_{r}/a)}  \ e^{\frac{\pi \omega_{r}}{a}} \non \\
&= &\frac{1}{2}\lbk 1+ e^{2\pi \omega_{r}/a}    \rbk.
\eqa
Utilizing the above expression, we find our emission rate in complete agreement with the Minkowski case, Eq. (\ref{mink}), when we identify the energy gap of the Unruh-DeWitt detector with the Rindler frequency, $\Delta E = \omega_{r} = k_{x}A\Omega$. Hence,
\bqa
\frac{d\Gamma}{d^{2}k_{\perp}} &=& -\frac{i \alpha}{4\pi a} 
 \lbk    \delta  H^{(2)}_{2i\omega_{r}/a}{\lb  \frac{-2ik_{\perp} \gamma }{a}\rb} -\frac{1}{2} \lb  H^{(2)}_{2i\omega_{r} +2}{\lb\frac{ -2ik_{\perp} \gamma }{a}\rb} +H^{(2)}_{2i\omega_{r} -2}{\lb \frac{-2ik_{\perp} \gamma }{a}\rb}	\rb \rbk \non \\
& \times & \lbk 1+ e^{2\pi \omega_{r}/a}    \rbk.
\eqa
\subsection{Acceleration Via Radiation Reaction}
Due to the similarity to channeling radiation, let us begin by presenting a simple example and compute the acceleration produced by radiative energy loss, i.e. bremsstrahlung. As such, we have

\bqa
\frac{dE}{dx} &=& - \frac{E}{x_{0}} \non \\
\Rightarrow E(x) &=& E_{0}e^{-x/x_{0}}.
\eqa 

For the sake of clarity, here we have $E = m\gamma = \sqrt{p^2 + m^2}$ and $p = mv\gamma = mu$. The parameter $x_{0}$ is the radiation length. Note, radiation lengths are typically on the order of $\sim $ cm, i.e. they are macroscopic quantities. As we will see, this parameter is what ultimately sets the length/time scale for the system and thus the acceleration as well. Let us consider an ultra-relativistic particle and assume the initial energy will be entirely due to the momentum $p_{0} = mu_{0}$, with $u_{0} = v_{0}\gamma_{0}$ being the initial proper velocity. Thus, the expression for radiative energy loss will reduce to the change in the relativistic momentum as a function of distance in the lab frame,

\bqe
p(x) = p_{0}e^{-x/x_{0}}.
\eqe

Note that this quantity determines the change in momentum of the positron. It is this change in momentum that gives rise to a force, and thus acceleration. Keeping all aspects of the analysis in the lab frame, let us recall the definition of force is the change in momentum per unit time, $f = \frac{dp}{dt}$. Thus 

\bqa
 \frac{dp}{dt} &=& ma' \non \\
 \Rightarrow a' &=& \frac{1}{m}\frac{dp}{dt}.
\eqa

Here, we explicitly have the proper acceleration, $a'$, from the relativistic version of Newtons law $f = ma_{lab}\gamma^{3}$. Again in the ultra relativistic limit, we obtain

\bqa
a' &=& \frac{1}{m x_{0}}\frac{dx}{dt}p_{0}e^{-x/x_{0}} \non \\
a'_{0} &=& \frac{\gamma_{0}}{x_{0}}.
\eqa 

Note that here we made use of the fact that $\frac{dx}{dt} = v = 1$ for relativistic velocities. Also, the approximation $e^{-x/x_{0}} = .96 \sim 1$ is equivalent to the time independence of the acceleration for the system. Note, using the parameters of the experiment; $\gamma = 3.5 \times 10^{5}$ and $x_{0}=9.37$ cm for silicon, the proper acceleration is given by $a'_{0} = 74$ ceV. Although this acceleration scale is relatively large in the Unruh setting, we need to find an acceleration scale at or beyond $\sim 100$ GeV, based on the energy scales in the experiment. Note again, the acceleration time scale is set by $t = x_{0}/c$ and therefore reflects the bulk acceleration produced by radiative energy loss. In order to bring about a larger acceleration we must look at the photon emission microscopically so as to obtain a more accurate acceleration time scale for each process individually.\\

From the power spectrum of the actual data set, we have a max photon frequency of $\omega_{0} = 150$ GeV. Note, this was also the first photon frequency to thermalize. Let us then examine the acceleration produced via this emission. Using this frequency as the change in momentum, $|\Delta p| = |k| = \omega_{0}$. This momentum change occurs during the lifetime of the emission process. Taking the physical size of the photon to be $\Delta x = \lambda/2$, we can then determine the emission time to be, $\Delta t = \Delta x/c = \frac{ \pi}{\omega_{0}}$. Then, using $\frac{\Delta p}{\Delta t} = ma'$, our proper acceleration is given by,

\bqa
a' = \frac{\omega_{0}^{2}}{ \pi m}
\eqa 

This is a proper acceleration but it is written in terms of the lab frequency. What is important to note is that when written in terms of proper quantities, the proper acceleration, $a' =\frac{\omega'^{2}_{0} \gamma^{2}}{ \pi m}$, boosts as $\gamma^{2}$. More importantly, when computing the FDU temperature for the emission, we find a recoil/radiation reaction temperature, $T_{RR}$. Hence,

\bqe
T_{RR} = \frac{\omega'^{2}_{0} \gamma^{2}}{ 2 m \pi^{2} }.
\eqe

Note, we now have an FDU temperature which depends explicitly on the recoil kinetic energy, $\omega^{2}/2m$ which is imparted on the positron by the emission. It is this acceleration, produced by the radiation reaction itself, that we need to look at in order to obtain large enough accelerations. Lets examine the temperature produced by the recoil of the maximum frequency in the data set, $\omega_{0} \sim 150$ GeV. As such, we expect to find a temperature of,

\bqa
T_{RR} & =& 2.23 \; PeV 
\eqa

\subsection{Bekenstein-Hawking Area-Entropy}

In consideration of the first law of thermodynamics, $d\mathcal{E} = T dS$, we can use this relation to determine the amount of entropy via the photon emission, here $\mathcal{E}$ is the energy radiated away in the proper frame. Note, all quantities here are necessarily in the proper frame. Recalling first that the temperature is given by $T_{RR} =\frac{\omega_{0}^{'2}\gamma^{2}}{2m \pi^{2}}$. From here we find that the temperature can be written in terms of the energy via $T_{RR} =\frac{\omega_{0}^{'2}E^{2}}{2m^{3} \pi^{2}}$ if we recall that $\gamma = E/m$. Note we express all our proper quantities in terms of the lab energy $E$. Then, recalling the proper energy is related to the lab energy via $d\mathcal{E} = dE/\gamma$, we have
\bqa
dS &=&\frac{2m\pi^{2} }{\omega_{0}^{'2} \gamma^{2}} d\mathcal{E} \non \\
&=&\frac{2m^4\pi^{2}}{\omega_{0}^{'2} } \frac{dE}{E^{3}} \non \\
\Rightarrow S &=& \frac{c^{8}}{\hbar^{2}}\frac{m^4\pi^{2}}{\omega_{0}^{'2} } \frac{1}{E^{2}}.
\eqa
Here, we assumed the initial entropy is zero. Note this expression is consistent with black hole entropy, and the Bekenstein-Hawking area-entropy law, since we have $ST = \frac{m}{2}$. We will now utilize this expression to examine the difference between the initial and final state entropy. Writing it explicitly in terms of the initial and final energy, in order to compare it to the change in the horizon area, we find,
\bqa
\Delta S &=& \frac{c^{8}}{\hbar^{2}}\frac{m^4\pi^{2}}{\omega_{0}^{'2} } \lbk \frac{1}{E^{2}_{f}}-\frac{1}{E^{2}_{i}} \rbk \non \\
&=& \frac{c^{8}}{\hbar^{2}}\frac{m^4\pi^{2}}{\omega_{0}^{'2} } \lbk \frac{1}{(E_{i} - \Delta E)^{2}}-\frac{1}{E^{2}_{i}} \rbk.
\eqa
It is this entropy that we will compare with the change in Rindler horizon area in order to experimentally confirm the proportionality factor of $\frac{1}{4}$. Note, $E_{f} = E_{i}  -\Delta E$, with $\Delta E$ given by the energy radiated by the positron and is determined by the integral of the power spectrum over frequency and time. It is also this energy, when boosted into the proper frame, that we will use to determine the change in the area of the Rindler horizon. As such, the change in area, $\Delta A$, generated by a flux of energy momentum across a Rindler horizon is given by \cite{satz},
\bqa
\Delta A = 8 \pi G \int d^{2}y \int_{0}^{\infty} dv v T^{\m \nu}k_{\m}k_{\nu}. 
\eqa
Here, $y = (y_{1},y_{2})$, is the transverse area, $v = \frac{x + t}{2}$ determines the spacetime propagation of the light like vector $k^{\m} = (1,1,0,0)$ which characterizes the light rays which span the horizon, and $T^{\m \nu}$ is the energy momentum ten	sor of the matter/energy that crosses through the horizon. The total mass/energy emitted by the positron into the horizon will give us $T^{\m \nu}k_{\m}k_{\nu} = \frac{\Delta E}{\gamma c^2}\delta{(v-v_{0})}\delta^2{(y)}$. Here, $\frac{\Delta E}{\gamma}$ is the total mass/energy  emitted by the positron in the proper frame. If we assume the energy is emitted at $x = \frac{1}{a}$ and $t=0$, then this implies the positron is at zero velocity in the proper frame. The mass/energy will then cross the horizon at $x = \frac{1}{a}$ and $t = \frac{1}{a}$. As such we have, $v_{0} = \frac{1}{a}$. Therefore, the total change in the horizon area is given by
\bqa
\Delta A &=& \frac{8 \pi G \Delta E}{\gamma ac^{2}} \non \\
&=& \frac{G c^{5}}{\hbar}\frac{8 \pi^2  m^{4}\Delta E}{E^{3} \omega_{0}^{'2} }
\eqa
The Bekenstein-Hawking area-entropy law states that $\Delta A/\Delta S = 4 \ell^{2}_{p}$, where $\ell^{2}_{p}$ is the Planck area. Writing the ratio of the area change to the entropy change yields,
\bqa
\frac{\Delta A}{\Delta S} = \ell^{2}_{p}\frac{8 \Delta E}{E^{3}_{i}}\lbk \frac{1}{(E_{i} - \Delta E)^{2}} -\frac{1}{E^{2}_{i}} \rbk^{-1}.
\eqa 

Note, to lowest order in acceleration and energy change, the area-entropy ratio is indeed trivially satisfied. This is due to the fact that the entropy is obtained by direct integration of the 1st law, $dE = T dS$, which a priori satisfies the Bekenstein-Hawking condition $S = A/4$. When expanding the entropy to first order in $\Delta E = 0$, and forming the ratio, this yields the first order term which will always be satisfied. To see this, let us consider more terms in the expansion. Hence,

\bqa
\frac{\Delta A}{\Delta S} &=& \ell^{2}_{p}\frac{8 \Delta E}{E^{3}_{i}}\lbk \frac{1}{(E_{i} - \Delta E)^{2}} -\frac{1}{E^{2}_{i}} \rbk^{-1}  \non \\
&\sim & 4\ell^{2}_{p}\lbk 1  - \frac{3 \Delta E}{2E_{i}}+\frac{ \Delta E^{2}}{4E_{i}^{2}}+\frac{ \Delta E^{3}}{8E_{i}^{3}} + \cdots \rbk. \non
\eqa 

What we find is that the zeroth order term is always satisfied. However, what is required to confirm the presence of thermality is that all terms satisfy the relation $S = A/4$, even with $\Delta E \neq 0$. Conceptually what this means is that by the original integration of the 1st law, we fix $S_{i} = \frac{A_{i}}{4}$. This is the zeroth order ``initial condition" of the integration. Then, we must have $\Delta E$ evolve in such a way that the change in the area and entropy also obey $\Delta S = \frac{ \Delta A}{4}$. This is not always the case in fact. The $\Delta E$ that presents itself in the above expressions must come from a thermalized observable. In other words, if the integral of $\Delta E$ came from a spectra with, e.g., a particle resonance present, then the resultant $\Delta E$ will not satisfy the Bekenstein-Hawking relationship.

To better illustrate this point, we have deformed the power spectrum data set by adding in a Gaussian resonance. The was accomplished by scaling the y-component of each data point as follows;

\bqe
data_{y}[i] \rightarrow data_{y}[i]*\lb 1+\frac{\alpha}{\sigma \sqrt{2\pi}} e^{(-1/2((i-\mu)/\sigma)^{2}} \rb.
\eqe

For our purpose we used $\alpha = 30$, $\sigma = 7$ and $\m = 50 \; GeV$ and $\mu = 70 \; GeV$. All errors were kept the same. We then use this spectrum in the integral of $\Delta E$. The resultant area-entropy ratio diverges from the Bekenstein-Hawking relationship and therefore reflects the lack of thermality in the system, see figure 1 below for an illustration of this phenomena.  

\begin{figure}[H]
\centering  
\includegraphics[scale=.35]{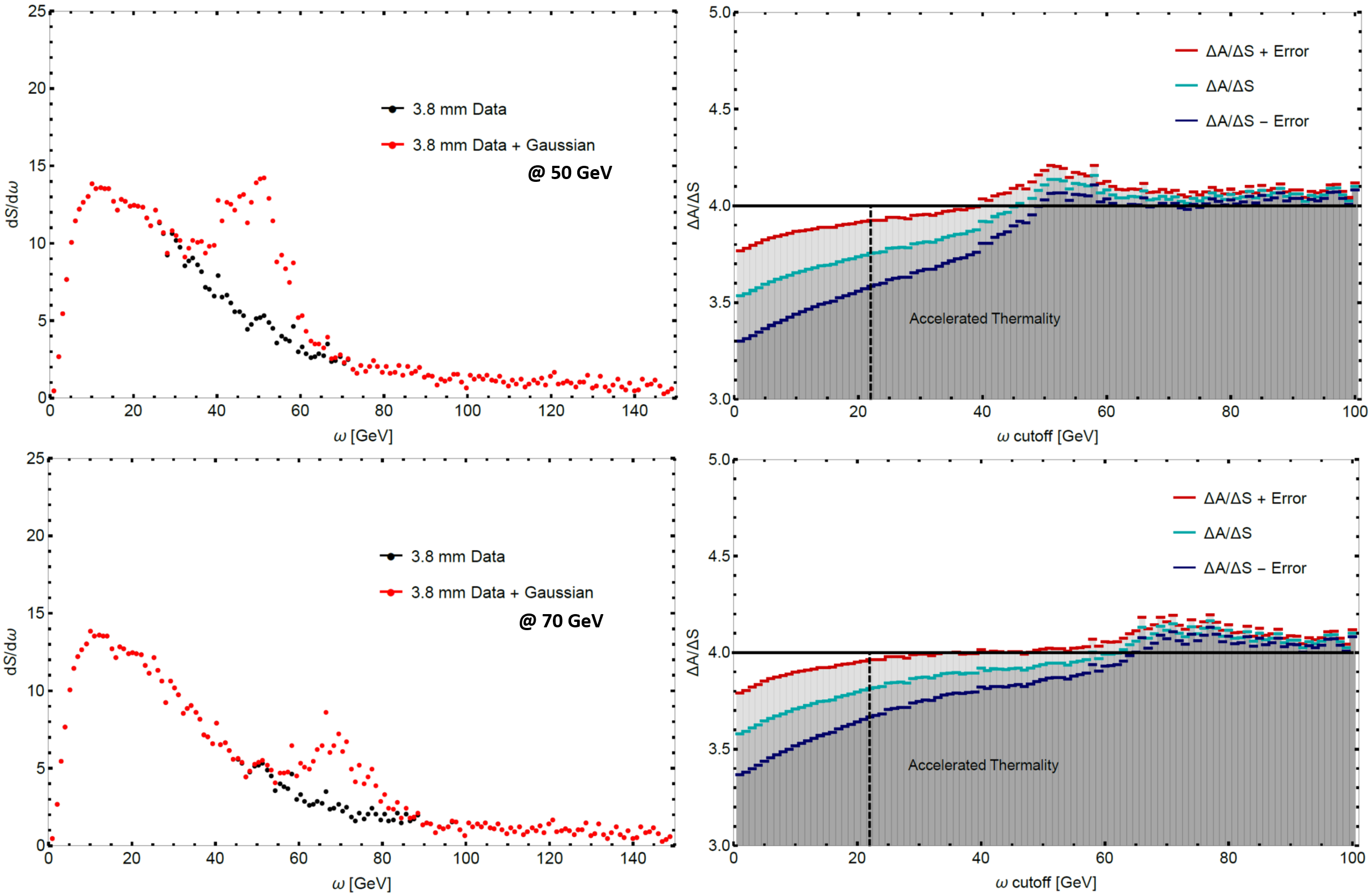} 
  \caption{\textbf{Area-entropy ratio for a non-thermal power spectrum:} By including a Gaussian resonance in the power spectrum data (LEFT), we can examine its effect on the area-entropy ratio (RIGHT). Note the Bekenstein-Hawking relationship fails for non-thermal power spectra simulated by including Gaussian resonances at $50 \; GeV$ (TOP) and $70 \; GeV$ (BOTTOM).}
\end{figure}

What the above figures demonstrate is that for the full dynamics over the integration of $\Delta E$, the area-entropy ratio will not be thermal if the observable analyzed is not thermal. In other words, satisfying the Bekenstein-Hawking relationsip $S = A/4$ must happen at all orders in $\Delta E$ and not just the initial condition, i.e. the initial condition satisfies $S_{i} = A_{i}/4$ and this relationship is upheld throughout the integration such that $\Delta S = \Delta A/4$, for all $\Delta E$ of a thermal system.\\

\subsection{Notes on the Energy Gap}

For the sake of simplicity, let us examine the radiation emission/absorption of a field $\phi(x)$ from a second quantized field $\psi(x)$, i.e. not an Unruh-DeWitt detector, in the Rindler frame. The integration over the spatial modes yields delta functions whose argument encodes conservation of momentum, e.g. $\delta(p_{f}-p_{i}+k)$, via	 

\bqa
\mathcal{A}_{i\rightarrow f} &\sim & \int d^{3}x \sqrt{-g} \psi^{\ast}_{f}(x,p)\psi_{f}(x,p)\phi(x,k) \non \\
&\sim & \delta(p_{f}-p_{i}+k). \non
\eqa

The expression for momentum conservation, in the case of the channeling experiment, can be rather complicated on account of there being a channeling oscillation, recoil, and all other processes present. When looking at the change in the electron Rindler energy $\Delta E = E_{f} - E_{i}$, we simply Taylor expand about the photon frequency. We must also note, that in an inertial comoving frame, the electron Rindler energy coincides with the electron Minkowski energy. Then, as an example, for the above conservation of momentum statement (taking $p_{i}=0$ for the sake of simplicity), we will have

\bqa
\Delta E &=& \sqrt{(p_{f})^2 +m^2} - E_{i} \\ \nonumber
&=& \sqrt{(-k)^2 +m^2} - m \\ \nonumber
&=& \sqrt{(\omega)^2 +m^2} -m \\ \nonumber
&\sim& \frac{\omega^2}{2m} \nonumber
\eqa 

We also expect there to be a pure channeling frequency $\Omega$ term based on both the Cozzella analysis \cite{matsas4} as well as the fact that the same term is present in the energy gap of an Unruh-DeWitt detector for the anomalous doppler effect. In our Rindler analysis, this term was most likely thrown away when we took the electron current to be at constant transverse velocity. However, by keeping the oscillation in the phase, we obtain the $\Delta E = \omega_{r} = k_{x}A\Omega$. For a dipole oscillator, when heavily boosted, this becomes beamed yielding $k_{x}\sim \omega$, in the lab frame. This yields our linear term, $A\Omega \omega$. These three terms together comprise the fiducial terms of our energy gap,

\bqe
\Delta E \sim \Omega + \Omega A \omega +\frac{\omega^{2}}{2m}.
\eqe

Finally, since we do not know what the exact dispersion relation is in the channeling experiment, we employed a more general polynomial in the emitted photon's frequency, i.e. $\Delta E = a_{0} + a_{1}\omega+ a_{2}\omega^{2}+ a_{3}\omega^{3}$. We include one more term beyond the known ones for the sake of completeness. To match the calculated spectrum to the data we also include an over all scaling factor $s$ for the spectrum and parameter, $\tilde{a}$, to fit the acceleration. The measured best fit parameters are presented below.

\begin{table}[H]
\centering
\begin{tabular}{cccccccc}
\hline \hline \\ [-2.0 ex]
Energy~&~$\chi^{2}/\nu$ & s ~~&~ $\tilde{a}$ [PeV] ~&~ $a_{0}$ [GeV] ~&~ $a_{1}$ ~&~ $a_{2}$ [GeV$^{-1}$] ~&~ $a_{3}$ [GeV$^{-2}$] \\
\hline \\ [-2.0 ex]
30 GeV &  $\;\;$ 1.114  $\;\;$& 12.93 $\;\;$&	7.939 & -0.00197 & 0.0120 & -846.2 & 0.4691 \\
\hline \\ [-2.0 ex]
\hline
\end{tabular}
\caption{The best fit parameters for our theoretical power spectrum with the energy gap for the 3.8 mm channeling crystal sample at the energy where the chi-squared statistic first meets the 1 standard deviation criterion. Our reduced chi-squared statistics shows that the data can be rigorously fit by the power spectrum with the energy gap thermalized at the FDU temperature.}
\end{table}

\subsection{Notes on the Thermalization Time}

Lets go back and consider an Unruh-DeWitt detector coupled to a massless scalar field. To begin, let us recall the response function, i.e. transition rate for the Unruh-DeWitt detector \cite{lynch}, is given by,

\bqa
\Gamma &=& q^{2} \int d\xi e^{-i\Delta E \xi} G^{\pm}[x',x].
\eqa

We should note here that the Wightman function $G^{\pm}[x',x]$ of the emitted scalar field, when evaluated on a hyperbolic trajectory, is given by $G = - \frac{1}{(2\pi)^{2}}\frac{a^{2}}{\sinh^{2}{(a \tau/2)}}$. The resultant integration yields the standard thermal response function associated with the Unruh effect,

\bqa
\Gamma &=& q^{2}\frac{\Delta E}{2\pi}\frac{1}{e^{2\pi \Delta E/a}-1}.
\eqa

The inverse of this expression is the thermalization time. However, we must note that the above expression utilized the Wightman function $G \sim 1/\sinh^{2}{(a \tau/2)}$, which had already integrated out the scalar fields frequency. To see this, let us examine the computation of the Wightman function explicitly, 

\bqa
G^{\pm}[x',x] &=& \bra{0}\hat{\phi}(x')\hat{\phi}(x)\ket{0} \nonumber \\
&=& \frac{1}{2(2 \pi)^{3}}\iint\frac{ d^{3}k'd^{3}k}{\sqrt{\omega'\omega}}\bra{0}\lbk \hat{a}_{\mathbf{k'}}e^{i(\mathbf{k'}\cdot \mathbf{x'} -\omega' t')}  + \hat{a}_{\mathbf{k'}}^{\dagger}e^{-i(\mathbf{k'}\cdot \mathbf{x'} -\omega' t')}\rbk\lbk \hat{a}_{\mathbf{k}}e^{i(\mathbf{k}\cdot \mathbf{x} -\omega t)}  + \hat{a}_{\mathbf{k}}^{\dagger}e^{-i(\mathbf{k}\cdot \mathbf{x} -\omega t)}\rbk\ket{0} \nonumber \\
&=& \frac{1}{2(2 \pi)^{3}}\iint\frac{ d^{3}k'd^{3}k}{\sqrt{\omega'\omega}}\bra{0} \hat{a}_{\mathbf{k'}}\hat{a}_{\mathbf{k}}^{\dagger}e^{i(\mathbf{k'}\cdot \mathbf{x'} -\mathbf{k}\cdot \mathbf{x} -\omega' t'+\omega t)}\ket{0} \nonumber \\
&=& \frac{1}{2(2 \pi)^{3}}\iint\frac{ d^{3}k'd^{3}k}{\sqrt{\omega'\omega}}e^{i(\mathbf{k'}\cdot \mathbf{x'} -\mathbf{k}\cdot \mathbf{x} -\omega' t'+\omega t)}\delta(\mathbf{k'}-\mathbf{k}) \nonumber \\
&=& \frac{1}{2(2 \pi)^{3}}\int\frac{d^{3}k}{\omega}e^{i(\mathbf{k}\cdot \Delta\mathbf{x} -\omega \Delta t)}.
\eqa   

Note, the frequency here is \textit{precisely} the frequency of the emitted particle; in this case a massless scalar but in our manuscript it is the photon. Integration of this expression over the frequencies yields the typical $G \sim 1/\sinh^{2}{(a \tau/2)}$ form and everything reduces to the standard Unruh case. Utilizing the unitegrated expression, the response function, and thus the basis for the power spectrum derived in the manuscript is given by,
 
\bqa
\Gamma &=&\frac{ q^{2}}{2(2 \pi)^{3}} \int d\xi  \int\frac{d^{3}k}{\omega} e^{-i\Delta E \xi}e^{i(\mathbf{k}\cdot \Delta\mathbf{x} -\omega \Delta t)}.
\eqa    
 
Letting $\Delta x = 0$, which is the ``non relativistic" approximation used in the manuscript that successfully yielded the Larmor formula, and recalling $\Delta t = \frac{2}{a}\sinh{(a\xi/2)\gamma}$ \cite{lynch1}, we then have

\bqa
\Gamma &=&\frac{ q^{2}}{2(2 \pi)^{3}} \int d\xi  \int\frac{d^{3}k}{\omega} e^{-i\Delta E \xi}e^{-i\omega \frac{2}{a}\sinh{(a\xi/2)\gamma}}.
\eqa 
 
Again, making the change of variables $w = a\xi/2$ and recalling the Hankel identity Eqn. (S40), we then have

\bqa
\Gamma &=&\frac{ q^{2}}{2(2 \pi)^{3}} \frac{2}{a}\int dw  \int\frac{d^{3}k}{\omega} e^{-i\Delta E \xi}e^{-i\omega \frac{2}{a}\sinh{(a\xi/2)\gamma}} \non \\
&=&\frac{ q^{2}}{(2 \pi)^{2}} \frac{2}{a}\int dw  \int d\omega \omega e^{-i\frac{2\Delta E}{a} w-i\omega \frac{2}{a}\sinh{(w)\gamma}} \non \\
&=&-i\alpha \frac{2}{a}\int d\omega \omega H^{(2)}_{\frac{2i\Delta E}{a}}\lb -i \frac{2\omega \gamma}{a}  \rb.
\eqa 

This is, of course, the analogous expression for the AQED emission rate, Eqn. (S41) along with $\frac{d\mathcal{S}}{d\omega} = \frac{d\Gamma}{d\omega}\omega$, but for the massless scalar field rather than photon; the main difference being the two additional Hankel functions which come from the polarization of the photon. Here, we see explicitly why we still have to integrate over the frequency. It is because we must start with the Wightman function which has not integrated the frequency out. This also enables us to include higher order frequency terms in the energy gap of the Unruh-DeWitt detector in a self consistent way. This essentially means that we have switched the order of integration between the proper time of the detector and the emitted photons frequency.

In terms of how to interpret the thermalization time, we must first comment on the above expressions relationship to the standard Unruh response function. Modulo the approximation $\Delta x = 0$, the integration over the frequency should reduce to the appropriate approximation of the standard Unruh form           	;

\bqa
\Gamma &=&-i\alpha \frac{2}{a}\int d\omega \omega H^{(2)}_{\frac{2i\Delta E}{a}}\lb -i \frac{2\omega \gamma}{a} \rb \non \\
&\sim & q^{2}\frac{\Delta E}{2\pi}\frac{1}{e^{2\pi \Delta E/a}-1}.
\eqa

This, of course, is valid provided switching the proper time and frequency integrals is valid. This quantity is to be interpreted as a decay/excitation rate in the conventional sense. In other words, given an initial population $N_{0}$ of excited detectors, the population as a function of time will be $N(t) = N_{0}e^{-\Gamma t}$. The thermalization time is given by $\tau = 1/\Gamma$ and determines the time until the population has been reduced by $1/e$. In the case of particle decays, this is known as the particle lifetime. This is also the quantity that is used as the time necessary for thermality to take hold in the Unruh setting \cite{matsas1}. Now, we note that the thermalization time is simply the unintegrated form, 

\bqe
\tau = \frac{1}{\int_{0}^{\infty}  \frac{d\Gamma}{d\omega} d\omega}.
\eqe

So the question is now, how does all of this change with a frequency dependent energy gap? The easiest way to see this is to look to an example of a particle decay with 3 branching ratios. Each process has a rate $\Gamma_{1}$, $\Gamma_{2}$, and $\Gamma_{3}$. The total decay rate is the sum off all three terms $\Gamma_{tot} = \sum \Gamma_{i}$. It is this total decay rate which shows up in the exponential decay, $N(t) = N_{0}e^{-\Gamma_{tot}t}$. The decay lifetime, or thermalization time, is the reciprocal of the sum, $\tau = 1/\sum \Gamma_{i}$, and not the sum of reciprocals, $\tau \neq 1/\Gamma_{1}+1/\Gamma_{2}+1/\Gamma_{3}$. Most importantly, each decay pathway has its own decay rate and will decay according to its own lifetime. The total decay rate simply combines all contributions for an ensemble system. The example here demonstrates what happens with discrete decay pathways. 

What about having an infinite number of decay/excitation pathways? Consider a ``particle" with an infinite number of Unruh-DeWitt detectors and a continuous distribution of energy gaps. With no degeneracies, we will have precisely one Unruh-DeWitt detector for every single frequency in the electromagnetic spectrum. This means the total decay rate will be summing over each mode $\Gamma_{t} = \sum_{i=0}^{\infty}\Gamma(id\omega)$. Since we have infinitesimal differences between our energy gaps, $d\omega$, then we should likewise have infinitesimal differences in our excitation rate, $d\Gamma(id\omega) = \Gamma((i+1)d\omega)-\Gamma(id\omega) $. Starting from the excitation rate of the zero energy gap, $\Gamma_{0}$, our sum then becomes $\Gamma_{t} = \Gamma_{0} + \sum_{i=0}^{\infty}d\Gamma(id\omega)$. For excitations, a zero energy gap will never excite and the rate is identically zero, $\Gamma_{0}= 0$. In the sum, we can multiply and divide by $d\omega$ to convert our expression into a Riemann sum, $\Gamma_{t} = \sum_{i=0}^{\infty}\frac{d\Gamma(id\omega)}{d\omega}d\omega$. This, of course, can now be transformed into an integral over the photon frequency and completes the derivation of the standard excitation rate for a continuous system as an integral over all internal energy gaps,

\bqe
 \Gamma_{t} = \int_{0}^{\infty}\frac{d\Gamma(\omega)}{d\omega}d\omega.  
\eqe

Let us now turn to the question of the thermalization time of an individual frequency. Recalling from the discrete case that the total decay rate was the sum of each rate of the individual decay modes we must also point at that each decay mode stands on its own. In other words, each individual decay pathway obeys its own decay rate, but the total decay rate depends on the sum. In just the same way, the individual excitation rate of a particular mode in a continuous system will also stand alone, and thermalize at its own prescribed time in the Unruh picture. In order to determine the excitation rate of an individual mode, $\omega'$ we must find the excitation rate of that frequency, i.e. $\Gamma(\omega')$, we must integrate up to that frequency in our total excitation rate. Thus,

\bqa
\int_{0}^{\omega'}\frac{d\Gamma(\omega)}{d\omega}d\omega  &=& \Gamma(\omega') - \Gamma(0) \non \\
 &=& \Gamma(\omega')
\eqa
Here we made use of the fundamental theorem of calculus and the fact that the excitation rate of a zero energy gap is zero. What this shows is that as we integrate over the frequency of a continuous detector energy gap, each frequency, $\omega'$, has an excitation rate, $\Gamma(\omega') = \int_{i=0}^{\omega'}\frac{d\Gamma(\omega)}{d\omega}d\omega$. The integral up to $\omega'$ gives precisely the individual excitation rate in the sum, of the total excitation rate, that characterizes this specific process individually. This process will also thermalize at its own time $\tau' = 1/\Gamma(\omega')$. This is precisely the scenario we saw in the discrete case. Each process thermalizes at its own time but the total thermalization time (meaning all processes have thermalized) is the inverse of the total rate.

\subsection{Notes on the Semiclassical Vector Current}
Let us go back and ``derive" the Unruh-DeWitt detector from the QED current interaction. This will show how it is possible to incorporate particle interactions into a two level system in a self consistent way. For the photon field, $\hat{A}^{\m}(x)$, We will have the following action \cite{peskin},

\bqe
\hat{S}_{I} = q\int d^{4}x \hat{\bar{\psi}} \gamma_{\m} \hat{\psi}\hat{A}^{\m}(x).
\eqe

Here we have the Dirac current, $\hat{j}_{\m} = \hat{\bar{\psi}} \gamma_{\m} \hat{\psi}$, which we would like to model as a semi-classical vector current. To see how this is accomplished we begin by recalling that for spinors $u(s,p)$ and $v(s,p)$ of spin $s$ and momentum $p$ that are created by $\hat{a}^{\dagger}_{s,p}$ and $\hat{b}^{\dagger}_{s,p}$, we have the following electron field operators

\bqa
\hat{\psi}(x,t) &=& \int d^3 p \sum_{s}\lbk \hat{a}_{s,p}u(s,p)\phi_{p}(x,t) + \hat{b}^{\dagger}_{s,p} v(s,p)\chi_{p}(x,t) \rbk , \non \\
\hat{\bar{\psi}}(x,t) &=& \int d^3 p \sum_{s}\lbk \hat{a}^{\dagger}_{s,p}\bar{u}(s,p)\phi^{\ast}_{p}(x,t) + \hat{b}_{s,p} \bar{v}(s,p)\chi^{\ast}_{p}(x,t) \rbk.
\eqa

The positive and negative frequency modes are given by $\phi_{p}(x,t)$ and $\chi_{p}(x,t)$ respectively. Normally these modes are plane waves, however in more general spacetimes our only requirement is that they are positive and negative frequency modes with respect to the particle's/detector's proper time. Using these fields, we will formulate the Dirac current, $\hat{j}_{\m} = \hat{\bar{\psi}} \gamma_{\m} \hat{\psi}$. Let us now consider the transition element between initial, $\ket{E_{i}}$, and final, $\ket{E_{f}}$, electron energy state. We are also neglecting any spin effects. Now, focusing strictly on electrons, i.e. no antiparticles, the only surviving element of the current will be given by, 

\bqa
\hat{j}_{\m}(x) = \bar{u}(p_{f})\gamma_{\m}u(p_{i})\phi^{\ast}_{E_{f}}(x,t)\hat{a}^{\dagger}_{p_{f}}\hat{a}_{p}\phi_{E_{i}}(x,t).
\eqa

In the above expression, we still have the spinor degrees of freedom to deal with. For this, we will make use of the Gordon identity,

\bqa
\bar{u}(p_{f})\gamma^{\m}u(p_{i}) = \bar{u}(p_{f})\lbk\frac{ p_{f}^{\m} + p_{i}^{\m} +i\sigma^{\m \nu}( p_{f\nu}- p_{i\nu})}{2m}\rbk u(p_{i}).
\eqa

Neglecting the spin coupling yields the kinematic term,

\bqa
\bar{u}(p_{f})\gamma^{\m}u(p_{i}) = \bar{u}(p_{f})\lbk\frac{ p_{f}^{\m} + p_{i}^{\m}}{2m}\rbk u(p_{i}).
\eqa

Finally, we will make the assumption that at the level of the semiclassical vector current, that the momentum remains constant throughout the radiative process. As such, we have 

\bqe
\bar{u}(p_{f})\gamma_{\m}u(p_{i}) = u_{\m}.
\eqe

We now make use of the fact that our positive and negative frequency mode solutions can be separated into their spatial and temporal components via $\phi(x,\tau) = g(x)e^{-iE\tau}$. We have chosen to parametrize our fields via the electron's proper time, $\tau$, to incorporate the Rindler coordinate chart when analyzing the accelerated case. Our current now reduces to

\bqa
\hat{j}_{\m}(x) = u_{\m}g_{f}^{\ast}(x)g_{i}(x)e^{i E_{f} \tau}\hat{a}^{\dagger}_{p_{f}}\hat{a}_{p_{i}}e^{-i E_{i} \tau}.
\label{sdelta}
\eqa

We note that for sufficiently localized electronic wave functions, e.g. with a wavelength much smaller than the wavelength of emitted radiation, we have $g_{f}^{\ast}(x)g_{i}(x) = \delta^{3}{(x - x_{tr})}$ along the classical trajectory of the electron; which is assumed to be uniform. Finally, by attaching the time-dependence to the creation and annihilation operators we have

\bqa
 e^{iE_{f}\tau}\hat{a}^{\dagger}_{p_{f}}\hat{a}_{p_{i}}e^{-iE_{i}\tau} &=& e^{i\hat{H}\tau}\hat{m}(0)e^{-i\hat{H}\tau} = \hat{m}(\tau).
\label{smono}
\eqa

Here we defined the Heisenberg evolved monopole moment operator $\hat{m} = e^{i\hat{H}\tau}\hat{m}(0)e^{-i\hat{H} \tau}$ where $\hat{m}(0)$ is defined as $\hat{m}(0)\ket{E_{i}} = \ket{E_{f}}$ with $E_{i}$ and $E_{f}$ the initial energy and final energy of the electron moving along the trajectory, $x_{tr}$, of the current. The energy gap of the detector is $\Delta E = E_{f} -E_{i}$ and the system is normalized via $1 =\vert \bra{E_{f}} \hat{m}(0)\ket{E_{i}} \vert$. As such, we have transformed our fermionic current into a semi-classical charged current coupled to an Unruh-DeWitt detector,

\bqe
\hat{j}_{\m}(x) = u_{\m}\hat{m}(\tau)\delta^{3}(x-x_{tr}). 
\eqe

We have provided a proof that the Dirac current can be reduced to a semiclassial vector current coupled to an Unruh-DeWitt detector. The energy gap is defined by the difference between the initial and final electron energy with respect to the proper time of the charged particle. Note, that the energy gap of our Unruh-DeWitt detector is given by, $\Delta E = a_{0} + a_{1}\omega+ a_{2}\omega^{2}+ a_{3}\omega^{3}$. Since the monopole moment operator creates and/or annihilates states with definite momentum then we need to think of our semiclassical vector current coupled to a continuum of Unruh-DeWitt detectors; one for each frequency of our energy gap.

\end{document}